\documentclass[lettersize,journal]{IEEEtran}
\usepackage{amsmath,amsfonts}
\usepackage{algorithmic}
\usepackage{algorithm}
\usepackage{array}
\usepackage[caption=false,font=normalsize,labelfont=sf,textfont=sf,labelformat=simple]{subfig}
\usepackage{textcomp}
\usepackage{stfloats}
\usepackage{url}
\usepackage{verbatim}
\usepackage{graphicx}
\usepackage{cite}
\usepackage{xcolor}
\usepackage{booktabs}
\usepackage{tabularx}
\usepackage{array}
\usepackage{multirow}
\usepackage{makecell}

\usepackage{xurl}

\usepackage{amssymb}

\usepackage[colorlinks=true, linkcolor=blue]{hyperref}
\newcolumntype{Y}{>{\raggedright\arraybackslash}X}

\begin{document}

\title{When 5G MIMO Scaling Breaks:\\ Toward 6G Upper-Mid-Band Extreme MIMO }




\author{Kwang Soon Kim,~\IEEEmembership{Senior Member,~IEEE}, Jeonghun Park,~\IEEEmembership{Member,~IEEE}, Byung-Wook Min,~\IEEEmembership{Senior Member,~IEEE}, Kwanghoon Lee,~\IEEEmembership{Graduate Student Member,~IEEE}, Eui Whan Jin,~\IEEEmembership{Graduate Student Member,~IEEE}, Juntaek Han,~\IEEEmembership{Graduate Student Member,~IEEE}, Geonwoo Park,~\IEEEmembership{Graduate Student Member,~IEEE}, Jun-Seok Ko,~\IEEEmembership{Graduate Student Member,~IEEE}, Jungho Myung,~\IEEEmembership{Member,~IEEE}, Wooram Shin,~\IEEEmembership{Member,~IEEE}, Young-Jo Ko,~\IEEEmembership{Member,~IEEE}, and Chan-Byoung Chae,~\IEEEmembership{Fellow,~IEEE}\\ \vspace{1.5em} \textit{Invited Paper}

\thanks{This work was in part supported by an Institute of Information \& Communications Technology Planning \& Evaluation (IITP) grant funded by the Korean government (MSIT) (No. RS-2024-00397216, Development of the Upper-mid Band Extreme Massive MIMO (E-MIMO)).}

\thanks{K. S. Kim, J. Park, B.-W. Min are with the School of Electrical and Electronic Engineering, Yonsei University, Seoul, 03722 South Korea. G.~Park and C.-B. Chae are with the School of Integrated Technology, Yonsei University, Seoul, 03722 South Korea. E-mail:\{ks.kim, jhpark, bmin, kwanghoon.lee, jinian, jthan1218, gwpark20, kjas9542 cbchae\}@yonsei.ac.kr. J. Myung, W. Shin, and Y.-J. Ko are with the Electronics and Telecommunications Research Institute (ETRI), Daejeon, 34129 South Korea. E-mail: \{jmyung, w.shin, koyj\}@etri.re.kr. K. S. Kim and J. Park are co-first authors.}
\thanks{\emph{Corresponding author: Chan-Byoung Chae}.} 

\thanks{Manuscript received July 30 XX, 2026; revised Nov. XX, 2026.}}

\markboth{Journal of \LaTeX\ Class Files,~Vol.~14, No.~8, August~2026}%
{Shell \MakeLowercase{\textit{et al.}}: IEEE Wireless Communications}


\maketitle

\begin{abstract}
The upper-mid band, particularly the 7–8 GHz range within frequency range 3 (FR3), has emerged as a leading spectrum candidate for wide-area sixth-generation (6G) cellular networks. Its shorter wavelength enables hundreds of antenna elements to be integrated within the physical aperture of an existing 5G base-station panel. In principle, the resulting aperture gain can compensate for the increased path loss and enable extreme MIMO (E-MIMO) with 256 or more antenna ports while reusing current cell sites. In practice, however, simply scaling the 5G New Radio (NR) architecture from tens to hundreds of ports encounters fundamental system-level limitations. This paper identifies where 5G-style MIMO scaling breaks and develops a research roadmap for practical upper-mid-band E-MIMO. We first review the evolution of FR3 spectrum, its propagation and channel characteristics, and the emerging 6G system requirements. We then organize the principal challenges into four coupled areas: maintaining effective coverage across all physical channels and protocol states; implementing wideband, energy-efficient RF devices and radio units; developing new low-power array and beamforming architectures; and acquiring sufficiently refined channel state information with manageable sounding and feedback overhead. Representative system studies illustrate the coverage asymmetry between user-specific data transmission and common or channel-acquisition signals, as well as the spectral- and energy-efficiency tradeoffs among fully digital, hybrid, tri-hybrid, dynamic-metasurface, and fluid-antenna architectures. Finally, we discuss how distributed apertures, integrated sensing, AI-assisted channel acquisition, and environment-aware operation can transform fixed-aperture scaling into a deployable 6G E-MIMO architecture.

\end{abstract}


\section{Introduction}
\IEEEPARstart{E}{ach} generation of cellular networks has been defined by a large increase in capacity, and the sixth generation (6G) is no exception. Its targets, a peak data rate on the order of 100\,Gbps and a user-experienced rate near 1\,Gbps, demand roughly a tenfold gain over
5G~\cite{samsung2020vision,itu2160}. 
Such a demand definitely calls for new spectrum. 
The upper-mid band, and the 7--8\,GHz range within frequency range~3 (FR3) in
particular, sits at a sweet spot: wide enough to supply the bandwidth 6G needs, yet low enough to retain the coverage of existing
networks~\cite{bjornson2025gmimo,andrews2024takeshape}.
Over the past few years it has moved from one candidate among many to a first-class 6G deployment band across standardization, industry, and
academia~\cite{wrc23,rel20sid,qualcommGigaMIMO2026}.

The bandwidth available in FR3 is nonetheless limited, so the capacity targets cannot be met by spectrum alone and rest on a major advance in
multiple-input multiple-output (MIMO)~\cite{bjornson2025gmimo}. 
The enabling observation is that the shorter wavelength of FR3 allows several hundred antenna elements to be integrated into the same physical aperture used at 3.5\,GHz. The resulting array gain compensates for the higher path loss and preserves the coverage of existing 5G sites. This fixed-aperture scaling to 256 or more antenna ports, referred to as extreme MIMO (E-MIMO), is thus the central direction of 6G physical-layer design~\cite{bjornson2025gmimo,qualcommGigaMIMO2026}.

Scaling the array by an order of magnitude, however, is not a matter of extending the 5G NR design. The equal-aperture argument guarantees coverage only in principle, at the level of propagation and form factor. 
Carried into a working system, the direct extension of the 5G standard runs into several challenges at the E-MIMO scale~\cite{equalAperture2026,3gppWorkshop2025}. 
Much breaks as the array grows from tens to hundreds of ports, each in its own way. Consider  first the coverage limitation. 
The equal-aperture argument applies to the data channel, which forms a user-specific narrow beam and so uses the full aperture gain. The control and sounding channels use wide beams, get no aperture gain, and so break long before data does. 
The hardware breaks next. At 7\,GHz over a fourfold wider band, the data converters, the beamforming network, and above all the power amplifier all lose ground at once. 
The amplifier suffers most, falling in the technology gap between the sub-6\,GHz and millimeter-wave bands. 
Power breaks as well. 
As the antennas multiply, the digital processing draws more power than the transmission itself. Keeping every element on and processing every chain digitally, as 5G does, does not scale to hundreds of chains. 
If the additional radiating elements are mapped directly to active RF chains, the converter and digital processing power can exceed the transmission power. The fully-digital scaling of every radiating element to an always-on RF chain is therefore difficult to sustain at the E-MIMO scale. The channel state information breaks last. Its acquisition cost grows too large at hundreds of ports, and the estimate is easily stale.


This paper examines where direct 5G MIMO scaling becomes inefficient or infeasible and identifies the research directions needed for deployable upper-mid-band E-MIMO. Section~II reviews the FR3 spectrum, channel characteristics, and emerging 6G system requirements, and establishes the four coupled challenges on this basis. Table~\ref{tab:mimo_scaling_breakpoints} summarizes the associated scaling mismatches, from physical-channel coverage and active hardware to distributed deployment and CSI acquisition. Sections~III--VI then examine these breakpoints in turn. For each, we identify which quantity fails to scale, explain why the corresponding 5G design becomes inadequate at the E-MIMO scale, and illustrate a representative research direction.



\begin{figure*}[!t]

\centering
\includegraphics[width=\textwidth]{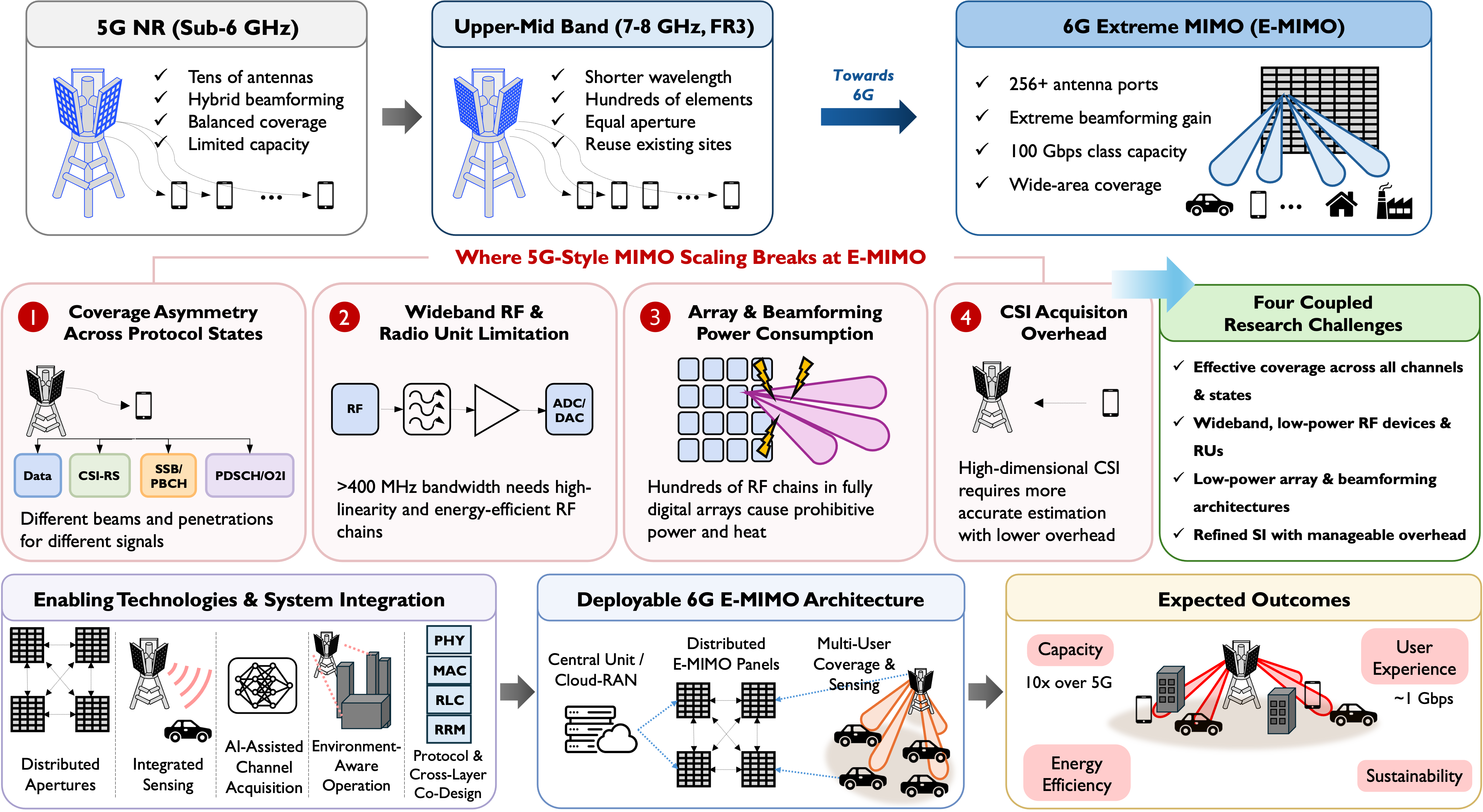}
\caption{Overview of the transition from 5G massive MIMO to deployable upper-mid-band extreme MIMO (E-MIMO). The shorter wavelength at 7--8~GHz enables hundreds of antenna elements to be integrated within an aperture comparable to that of an existing 5G base-station panel. A direct extension of the 5G MIMO architecture, however, encounters four coupled system-level limitations: coverage asymmetry across physical channels and protocol states, wideband and energy-efficient RF and radio-unit implementation, excessive array and beamforming power consumption, and rapidly increasing CSI-acquisition overhead. Addressing these limitations requires protocol-aware coverage enhancement, low-power RF and array architectures, distributed apertures, integrated sensing, AI-assisted channel acquisition, and environment-aware cross-layer operation.}
\label{fig:overview}
\end{figure*}

\section{6G Spectrum, Channel Characteristics, System Requirements, and Technical Challenges}

\subsection{6G Spectrum}

Since the bands below 6\,GHz are already saturated, the search for 6G spectrum
spanned every candidate band, from low-band through mid-band to millimeter-wave
and sub-THz. The upper-mid band (7--24\,GHz) was only one of these, with no clear standing in academia, industry, or standardization. The channel measurements and models for 6--24\,GHz were still a gap, leaving little basis to support its
use~\cite{wang2020channels}. Industry folded it into a broad all-band vision
centered on sub-THz, and standardization had opened no dedicated track for it. This changed in 2021. 3GPP opened a formal study item on 7--24\,GHz for
NR~\cite{tr38820}, and the centimeter-wave range was put forward as the 6G
spectrum element that could fill the gap between the sub-6\,GHz and
millimeter-wave bands~\cite{tongzhu2021}. In 2022, the band was named a 6G
candidate, balancing the limited bandwidth of sub-6\,GHz against the limited
reach of millimeter-wave~\cite{samsung2022spectrum}. Even then it was only one
option in a portfolio, still weighed against sub-THz and the re-farming of
existing bands. The turn came in 2023, when the band was elevated on all three
fronts at once. WRC-23 identified the upper-6\,GHz band (6.425--7.125\,GHz,
region by region) for IMT, and placed 7.125--8.4\,GHz, 14.8--15.35\,GHz, and
4.4--4.8\,GHz on the WRC-27 agenda as 6G study bands~\cite{wrc23}. A projected
shortfall of 1.5--2.2\,GHz of wide-area spectrum by 2030 marked the 7--15\,GHz
range as the best complement~\cite{ericsson2023spectrum,gsma2023}. And the band
was formalized as FR3, a wideband system that exploits spatial and frequency
degrees of freedom~\cite{kang2024fr3,cui2023fr3}. 
After 2024, the question turned from whether the band belongs in 6G to how it can be used, and on what technical basis. Academia and industry took up that question in parallel, the one through analysis and the other through prototypes.

On the academic side, it was shown in~\cite{bjornson2025gmimo} that the upper-mid-band
spectrum is too limited to reach the ITU peak-rate target on its own, so the shortfall
must be closed by scaling MIMO. 
At the same time, it was further shown that packing more antenna elements into the same physical panel, from about a hundred at 3.5\,GHz to several hundred at
7--8\,GHz, can compensate for the path loss that grows with
frequency~\cite{bjornson2025gmimo}. Configurations with at least 256 antenna ports
were referred to as gigantic MIMO. This shifted the basis for using the band from the
balance of its propagation characteristics to MIMO scaling under a fixed aperture. In
the same vein,~\cite{andrews2024takeshape} gave four reasons why the 7--8\,GHz band is
suited to the roughly tenfold capacity increase targeted for 6G. 1)~Its propagation
is adequate for cellular coverage and, unlike mmWave, allows outdoor-to-indoor
penetration. 2)~A modest increase of the array dimensions within the same panel can
provide coverage similar to current C-band deployments, so the existing cell sites
can be reused, which is decisive in terms of cost. 3)~The available bandwidth is
large, on par with all of the current sub-6\,GHz spectrum. 4)~The short wavelength
makes large arrays feasible, so considerable spatial multiplexing is expected. On the
industry side, Qualcomm presented Giga-MIMO prototypes that fit thousands of antenna
elements and hundreds of transmit-receive chains into the panel form factor already
used for the lower mid-band~\cite{qualcommGigaMIMO2026}. Narrower beams raise the
radiated power for the same conducted power, so the increase of the propagation loss
is compensated. This lets the upper-mid band serve as a wide-area 6G layer that is
co-located with the existing 5G mid-band sites. Nokia joined
the same direction under its own notion of extreme massive MIMO, through a coverage
evaluation of the 7--15\,GHz bands from existing
sites~\cite{nokiaExtreme2021,nokiaCoverage2024}. Huawei pursued the same direction with an upper-6\,GHz product line that builds on
an extremely large antenna array and targets coverage comparable to the
C-band~\cite{huaweiU6ghz2026}.

The claim that antenna scaling under a fixed aperture provides 5G-level coverage then
entered a phase of validation through measurements and standard channel models. The
dual-band measurements by NYU provided a propagation-physics basis, which showed that
a high-gain antenna of the same aperture can offer comparable coverage even at higher
frequencies~\cite{shakya2024indoor,shakya2024outdoor}. This basis rests on three
observations. 1)~In omnidirectional line-of-sight (LOS) conditions, the path-loss
exponent stays below two, that is, below the free-space exponent, owing to a
waveguiding effect. 2)~In directional non-line-of-sight (NLOS) conditions, the
path-loss exponent is lower than that of millimeter-wave. 3)~Long-range links are
established in outdoor urban-microcell (UMi) environments. The same measurements also
revealed that the directional NLOS path-loss exponent increases with frequency. These
measurement results were reflected in the path-loss models of 3GPP Release~19. The
white-paper-level claims were further extended to outdoor trials by Nokia, Ericsson,
and Qualcomm, and to prototypes by ZTE and
Samsung~\cite{poddar2025rel19,mwc25trials,qualcommGigaMIMO2026,zte2026gigamimo,samsung2026fieldtest}.
The path-loss compensation provided by the array gain of the same aperture was
sufficient to support the reuse of existing sites in LOS or light NLOS environments,
where a strong dominant path exists. As the environment moves toward deep NLOS,
however, a residual loss remains. In such conditions, blockage becomes severe and the
dominant path disappears~\cite{equalAperture2026,bazzi2025vision}.

The developments above converge into a consensus. FR3, and in particular
7.125--8.4\,GHz for the initial deployment, is taken as a first-class 6G deployment
band~\cite{rel20sid,3gppWorkshop2025,qualcommGigaMIMO2026,bjornson2025gmimo}. Since
the claim of obtaining the same coverage under the same aperture holds approximately,
the direction of 6G MIMO is set not by the amount of spectrum but by increasing the
number of ports and elements under an aperture constraint. The concrete MIMO design
for this direction must, however, account for the additional NLOS loss confirmed
above. It must also account for channel characteristics that cannot be reduced to the
path-loss exponent alone~\cite{equalAperture2026,bjornson2025gmimo}. The co-located
grid is a guideline, not a constraint. It comes from the convenience of reusing
existing sites, not from any claim of optimality. New sites are therefore also
deployed off the grid where capacity demands them, with priority-based cell
reselection required accordingly~\cite{lopezperez2025multilayer}.

\subsection{Upper-mid Band Channel Characteristics}

The frequency dependence of path loss, considering LOS, enters not the distance exponent but the free-space reference term. As the frequency increases, the effective aperture of a fixed-gain antenna shrinks, so this reference loss grows in proportion to
$f^{2}$~\cite{shakya2024indoor,bjornson2025gmimo}. The wideband measurements by NYU were conducted with horn antennas whose gain increases with frequency, that is, under a roughly fixed physical aperture. They reported both the directional path loss
including the horn gain and the omnidirectional path loss with the gain removed, which
separated the antenna effect from the channel itself~\cite{shakya2024indoor}. Under
these conditions, the LOS path-loss exponent stayed at or below about two across the
indoor, outdoor, and factory environments owing to a waveguiding effect, and it was
nearly independent of frequency~\cite{shakya2024indoor,ying2025inf,shakya2024outdoor}.

In NLOS, by contrast, the path-loss exponent increased with frequency, by about 17\%
and 30\% in the indoor and factory environments, respectively. Its value nevertheless
remained slightly lower than at the adjacent 28\,GHz mmWave band and much lower toward
the sub-THz range above 73\,GHz, so the NLOS advantage of FR3 stood out most clearly
against the higher bands. In the factory (InF), metal scattering compensated for the
blockage and kept the NLOS loss relatively lower. The outdoor (UMi) environment
followed the same trend and remained below mmWave and sub-THz primarily in
NLOS~\cite{shakya2024indoor,ying2025inf,shakya2024outdoor}. Taken together, the $f^{2}$
reference loss that grows with frequency is offset when the physical aperture is fixed
and the antenna gain is scaled as $f^{2}$. A high-gain antenna of the same aperture
therefore attains a coverage advantage at higher frequencies, and the
frequency-dependent additional loss in NLOS is what remains. This consensus was
reflected in Release~19, where the existing path-loss models including the indoor hotspot (InH) were retained as consistent with the FR3 measurements. 
For the building penetration loss in
urban environments (UMi, UMa), however, the measurements concluded that the
penetration loss of materials such as concrete and infrared-reflective glass (IRR glass) increases with frequency, so a revision of the existing models is possibly needed. Owing to the scarcity of measurement data in 7--24\,GHz, this re-parameterization has not yet been reflected in the standard~\cite{bjornson2025gmimo,poddar2025rel19,shakya2024indoor}.

In the same measurement series, it is reported that the RMS delay spread decreased with frequency across
the indoor, outdoor, and factory environments, which showed that the multipath becomes more concentrated in time at higher frequencies~\cite{shakya2024indoor,ying2025inf,shakya2024outdoor}. The angular spread likewise narrowed with frequency, and NLOS was wider than LOS. Across environments including the urban (UMi) case, the channel therefore became concentrated over a
narrower range in the angular domain as
well~\cite{shakya2024indoor,ying2025inf,shakya2024outdoor}. In other words, the FR3
channel has fewer multipath components gathered over a narrower range in both time and angle than the existing 3GPP models assume. This is favorable for beamforming that concentrates energy in a narrow direction, but it might limit the gain of spatial multiplexing that sends mutually independent streams at the same time~\cite{shakya2024indoor,ying2025inf,shakya2024outdoor}. Release~19 reflected this in
three ways. First, the delay-spread and angular-spread distribution parameters for UMi
and UMa were re-fitted to the new measurements. Second, the previously fixed number of
rays per cluster was changed into a frequency-dependent variable that can fall below
20. Third, a variable number of clusters per link was allowed. These changes represent the actual FR3 channel, which has fewer multipath components and a limited multiplexing gain~\cite{poddar2025rel19}.

This limit, however, holds only at a fixed array size, and equal-aperture scaling relaxes it.
The dual-band outdoor UMa measurements matched the same physical aperture across two bands that are roughly a factor of two apart. That is to say, the lower band used a 32-element array, and the higher band used a 128-element planar array. According to these measurements and ray-tracing analyses for indoor and outdoor environments, the angular spread
narrows as the frequency increases. The signals received by the elements therefore become more similar to one another (increased correlation). The channel eigenvalues also concentrate on a few
modes~\cite{shakya2024indoor,equalAperture2026,siteSpecific2026}. The resulting change in MIMO performance is split by environments. 
In indoor environments rich in multipath,
the number of independent streams that can be sent simultaneously (the rank) remains
high and decreases only gently with frequency. Outdoors, the shorter wavelength
resolves the paths better in angle, so the rank is maintained or even increases, and spatial multiplexing remains feasible even at higher frequencies~\cite{siteSpecific2026}.
Here the channel sparsity is compensated by placing four times as many elements within
the same aperture, since the number of elements scales as $1/\lambda^{2}$. The spectral
efficiency is in fact higher in the higher band. 
The two
orthogonal polarizations also remain well separated after passing through the channel,
which provides two nearly independent signal
paths~\cite{bjornson2025gmimo,equalAperture2026,shakya2024indoor}. 

Transplanting the 5G
64TR aperture directly to FR3, however, yields a 256TR scale. Once a 5G-level subarray
is taken into account, this reaches on the order of 768 to 1024 elements. The
covariance at this scale calls for further measurement and analysis, together with
whether the fading follows a Rician, a Rayleigh, or some other
distribution~\cite{equalAperture2026,siteSpecific2026,bjornson2025gmimo}. The aperture
may also grow beyond this level of 5G reuse, or the user may lie at a close range
within the Fraunhofer distance. In those cases, the near field (NF) and the spatial
non-stationarity (SNS) must be additionally considered. In the near field, the
wavefront is treated as spherical rather than planar. Under spatial non-stationarity,
only part of the array is blocked or scattered, so the received power varies across the
elements. These two effects were introduced into Release~19 as a spherical-wavefront
model and an element-wise power-variation model,
respectively~\cite{nfsns2025,poddar2025rel19}.

\subsection{Current Consensus on 6G System Requirements}

Before Framework 2030, the discussion of 6G system requirements rested on a loose,
quantitatively framed consensus. It assumed the use of the upper-mid band around
7\,GHz, the reuse of the sites and coverage of 5G, and a performance increase by a
factor of 5--10~\cite{samsung2020vision,qualcommGigaMIMO2026,3gppWorkshop2025,bjornson2025gmimo}.
In late 2023, the ITU-R IMT-2030 Framework (M.2160) kept the quantitative targets
conservative, with a peak data rate of 50--200\,Gbps, a user-experienced rate of
300--500\,Mbps, and a spectral efficiency 1.5--3 times that of IMT-2020. In addition
to the three scenarios evolved from 5G (i.e., eMBB, mMTC, and URLLC), it also introduced three new usage scenarios
that previous generations were not designed to support, namely AI and communication,
integrated sensing and communication (ISAC), and ubiquitous connectivity~\cite{itu2160}.
This shift suggests that 6G is not a mere quantitative improvement over 5G. Much as
data turned from an auxiliary of voice into the essence of the system in the transition
from the third to the fourth generation, the essential role of the mobile network
changes toward a core infrastructure that performs and connects distributed AI
computing~\cite{itu2160,andrews2024takeshape}. On the industry side, the AI-RAN vision
presented by NVIDIA concretizes this direction, and it recasts the RAN from a pure
connectivity layer into a distributed edge AI computing platform~\cite{nvidiaAiRan}.
This change of role is not yet a single agreed-upon requirement, but it is the
direction clearly indicated by the framework and by the converging industry vision.

Across the 2025 3GPP workshop and MWC 2026, the visions of operators and major vendors
converged on a common view. The essential goal of 6G is to make the RAN a computing
platform that cooperatively connects AI distributed across the device, edge, and
cloud~\cite{nvidiaAiRan,huaweiNet4ai,aiRanMWC26}. This is the realization of AI-on-RAN,
where AI inference and services run on top of the RAN. In particular, Huawei formalizes this most explicitly under its Connected Intelligence vision, as what it calls 6G for
AI~\cite{huaweiNet4ai,tongzhu2021}. In the NET4AI architecture, each network element
natively integrates communication, computing, and sensing. It also provides
large-scale distributed training and real-time edge inference as AI as a Service, which
shifts the center of gravity of intelligence from the central cloud to the network
edge. On the infrastructure side, NVIDIA's AI-on-RAN starts from the need for the same
RAN equipment to perform both radio signal processing and distributed edge AI inference
(e.g., vision AI agents)~\cite{nvidiaAiRan}. This transforms the existing RAN into an
edge AI computing platform. The vision is taking concrete shape through operators'
AI-RAN trials and hubs, and it is entering an industry-level demonstration phase.
Representative examples are T-Mobile, in collaboration with NVIDIA and Nokia, the
Deutsche Telekom 6G Innovation Hub, and Samsung, which promotes an AI-native
vRAN~\cite{aiRanMWC26,samsungAiRan}.

The RAN technology carrying this role must support 256 or more ports in the
standard, at bounded overhead and guaranteed performance. The conventional
approach relies on predetermined pilots, fixed modulation, and standardized
channel-state feedback, so the overhead grows rapidly with the port count. 
What is needed instead is an approach that layers AI (AI-for-RAN) and sensing
(ISAC) on top of radio-signal-based estimation, cutting the overhead while
raising the performance. Huawei's AI for 6G is
exactly this, a learning-based AI-native air interface with native
ISAC~\cite{huaweiNet4ai,tongzhu2021}, and Nokia's replaces predetermined pilots
and feedback with learned models~\cite{nokiaAiAirInterface}. Ericsson pursues
AI-based network automation on a 5G SA foundation, and Qualcomm an AI-based
channel-state feedback that gains even without site-specific
training~\cite{qualcommGigaMIMO2026,aiRanMWC26}. Release~20 makes both
foundations first-class for FR3: the extreme-scale MIMO above, and ISAC as a
Day-1 capability.

Ubiquitous connectivity extends 6G beyond the civilian services and sensing of
commercial networks. Rather than a settled requirement, it is a directional
consensus that new functions and roles are needed~\cite{itu2160}, and it takes
three forms. The first is connectivity where commercial coverage does not reach,
such as deserts, mountains, and oceans, carried mainly by satellites and
non-terrestrial networks (NTNs)~\cite{itu2160,ntnImt2030,rel20sid}. The second is
high-quality connectivity aboard ships and
aircraft~\cite{maritime6g,ntnImt2030}. The third is sensing and connectivity for
public-safety and defense use, where coverage must be raised at a chosen time and
place that no commercial network
serves~\cite{nvidiaAiRanStack2025,publicSafetyMC}. Realizing this expansion calls
for advances in non-terrestrial integration and mission-critical operation.

\subsection{Remaining Technical Challenges}

The directional consensus established in the preceding subsection guarantees only an
in-principle possibility, at the level of propagation and form factor. It states that
coverage can be recovered even in the 7\,GHz band by reusing 5G sites with an equal
aperture. When this is carried into an actual system, however, simply extending the 5G
NR standard technology immediately runs into a number of technical
walls~\cite{equalAperture2026,bjornson2025gmimo,3gppWorkshop2025}. These walls
ultimately converge into a single coupled problem. The promise of coverage,
5G-equivalent or better at every location and on every channel, must be kept, while power consumption and the cost of acquiring channel state information (CSI) keep growing~\cite{lopezperez2025multilayer}. 
This section organizes the walls into four categories, each of which is further elaborated into the subsections that follow. The four categories are securing effective coverage throughout the whole protocol, the
implementation of wideband and low-power RF devices and radio units (RUs), new structures and
devices that drastically reduce power, and CSI acquisition that is at once more refined
and lower in overhead.

For help of understanding, we summarize the scaling mismatches underlying these four categories in Table~\ref{tab:mimo_scaling_breakpoints}. 
The rows do not correspond one-to-one to the four challenge categories. Rather, they identify the physical, hardware, and protocol quantities whose coupled scaling gives rise to those challenges. The entries should not be interpreted as normative 6G requirements; they represent the implementation scales that arise when a 5G massive-MIMO panel is extended to upper-mid-band E-MIMO under a comparable physical aperture.


\begin{table*}[t]
\centering
\caption{Representative scaling from 5G massive MIMO to upper-mid-band
6G E-MIMO and the associated system-level bottlenecks. The values indicate
representative implementation scales rather than normative 6G requirements.}
\label{tab:mimo_scaling_breakpoints}
\footnotesize
\setlength{\tabcolsep}{4.0pt}
\renewcommand{\arraystretch}{1.20}
\begin{tabularx}{\textwidth}{
    >{\raggedright\arraybackslash}p{0.16\textwidth}
    >{\raggedright\arraybackslash}p{0.105\textwidth}
    >{\raggedright\arraybackslash}p{0.125\textwidth}
    Y
    Y
}
\toprule
\textbf{Scaling quantity}
&
\textbf{Representative 5G baseline}
&
\textbf{Representative 6G regime}
&
\textbf{What breaks under direct scaling}
&
\textbf{Required direction}
\\
\midrule

Physical-channel coverage (Section III)
&
5G-scale beam hierarchy
&
Narrower and more numerous beams
&
Common, control, and acquisition channels cannot fully exploit
the UE-specific aperture gain
&
Protocol-aware hierarchical beam management
\\

Radiating elements per panel (Section IV)
&
64--256
&
512--1,024+
&
Feeding-network loss, mutual coupling, thermal density, and calibration complexity
&
Reconfigurable and architecture-aware apertures
\\

\addlinespace[2pt]
Antenna ports (Sections~IV-V)
&
32--64
&
256+ logical ports
&
CSI-RS sounding and CSI-feedback overhead
&
Effective-rank channel acquisition
\\

\addlinespace[2pt]
Active RF chains (Section IV)
&
32--64
&
256+ under direct scaling
&
Data-converter, RF-chain, and baseband power consumption
&
Hybrid/tri-hybrid processing and adaptive RF-chain activation
\\

\addlinespace[2pt]
Instantaneous bandwidth (Section IV)
&
Up to 100 MHz
&
400 MHz+
&
PA efficiency, beam squint, and DPD/CFR complexity
&
Wideband RF--antenna--baseband co-design
\\

\addlinespace[2pt]
Aperture distribution (Section V)
&
Primarily co-located
&
Distributed or heterogeneous
&
Fronthaul, synchronization, calibration, and coordination overhead
&
User-centric distributed MIMO with selective RU activation
\\

\bottomrule
\end{tabularx}
\end{table*}

It is worthwhile to note that the desired 6G evolution is not a uniform enlargement of every dimension of a 5G radio. The accessible radiating aperture may grow to hundreds or thousands of elements, while the number of active RF chains and the CSI-acquisition dimension should follow only the spatial degrees of freedom required by the propagation environment and the served users. Likewise, distributing the aperture can reduce propagation loss and blockage, but only when the resulting fronthaul, synchronization, calibration, and static power costs are controlled. 
In what follows, we examine these coupled breakpoints with reference to Table~\ref{tab:mimo_scaling_breakpoints}.


The equal-aperture argument that the free-space path loss is compensated is a
proposition about the PDSCH. The PDSCH fully exploits the aperture gain through a
UE-specific narrow beam~\cite{equalAperture2026,bjornson2025gmimo}. Whether a cell is
seen, acquired, and managed, however, is determined by the link budgets of the
wide-beam common signals (SSB), the estimation signals (CSI-RS/SRS), and the uplink.
None of these enjoy the narrow-beam aperture gain in the same way~\cite{nr38214}.
The coverage of 7\,GHz E-MIMO is therefore a problem of structural
asymmetry. Even though the data may in principle reach the user, cell search, beam
management, and CSI acquisition break first. Resolving this asymmetry is what makes
coverage 5G-equivalent or better at every location and on every channel.
\begin{itemize}
  \item \textbf{Efficient, equivalent coverage across the whole cell}: at every location
  in the cell, a quality equal to or better than that
  of 5G must be secured efficiently. This must be done on the basis of an understanding
  of the propagation environment, rather than through indiscriminate reinforcement such
  as raising the transmit power~\cite{equalAperture2026}.

  \item \textbf{Equivalence across all channels of the protocol}: the various channels
  required by the protocol, namely SSB, CSI-RS, PDCCH, and the uplink channels, do not
  enjoy the narrow-beam aperture gain. This asymmetry must be closed through hierarchical
  beam search, a multi-band anchor, uplink enhancement, and reliable beam
  management~\cite{lopezperez2025multilayer}.

  \item \textbf{Restoration of coverage targeted especially at the indoor case}: the
  coverage is expected to deteriorate especially indoors, in the deep interior of
  high-rise buildings, owing to coated curtain walls and multiple-wall penetration. This
  coverage loss must be restored~\cite{shakya2024indoor,mwc25trials}.

  \item \textbf{Fast, adaptive configuration of variable coverage}: even under coverage
  that changes with the situation, as in a non-civilian dedicated network, effective
  coverage must be configured quickly and adaptively~\cite{publicSafetyMC}.
\end{itemize}

A 5G mid-band RU is generally implemented as a structure that combines digital
beamforming followed by subarray-level analog beamforming. Extending it to FR3, where
the frequency rises to 7\,GHz and the bandwidth grows roughly fourfold, within the same
aperture and form factor entails the following difficulties of RF-device and RU
implementation, together with the technical advances they require~\cite{cui2023fr3}.
\begin{itemize}
  \item \textbf{RF devices and transceivers integrated at high efficiency over a wide
  band at high frequency}: at 7\,GHz and beyond 400\,MHz, the ADC/DAC, the RF components
  of the beamforming network, and the PA all become simultaneously disadvantaged in
  conversion speed, insertion loss, efficiency, and integration density. The FR3 PA
  technology sits between the sub-6 LDMOS and the mmWave SiGe/CMOS regimes, so
  integrating devices at high efficiency within the same aperture is
  difficult~\cite{cui2023fr3}.

  \item \textbf{An analog beam network that varies at low power and high speed}: a fixed
  subarray beam cannot adapt the beam and the power to the per-site and per-location
  propagation environment and traffic. A network based on simple phase shifters has
  large insertion loss and power and a slow varying response. What is needed is RF and
  antenna technology, together with its calibration, that is more power-efficient than a
  phase-shifter network. This technology must rapidly change the beam direction and the
  per-PA output power, so as to provide equivalent coverage at low power at the same
  site~\cite{ttdTriHybrid,otaCalibration}.

  \item \textbf{Linearization that is low-power and high-quality over a wide band at high
  frequency}: satisfying both DPD/CFR and ACLR over an instantaneous bandwidth widened
  roughly fourfold and across 256 ports makes the oversampling of the feedback receiver
  and the per-port computation explode, so maintaining high linearity at low power is
  difficult~\cite{hybridDpd,mmimoLinearization}.

  \item \textbf{An RF and array that integrate communication and sensing}: ISAC requires
  that the RF provide radar-like functions, such as canceling the transmitter's own
  signal or dynamically and rapidly dividing transmission and reception in time and
  frequency. It also requires that both efficient beamforming over a fixed narrow region
  and radar-like sweeping over a wide region be supported in a variable
  manner~\cite{fullDuplexIsac,Masouros2022beyond6G}.
\end{itemize}

As the number of antennas grows, the digital processing power comes to exceed the transmit (PA) power by several times. A direct fully-digital extension in which the growing aperture
is supported by an equally large number of always-active RF
chains therefore cannot satisfy the power budget of a
256-TRX-scale array. The following new structures, devices, and signal processing,
beyond a simple extension, are therefore needed, even though they are hard to realize at
present~\cite{powerModelFr3}.
\begin{itemize}
  \item \textbf{A new distributed array and beamforming structure unlike the present
  one}: to reduce power drastically, the aperture should be partitioned among spatially
  separated RUs rather than concentrated at one site. Participating RUs should perform
  joint beamforming over the distributed aperture, with coherence used only when its gain
  justifies the synchronization overhead~\cite{bjornson2025gmimo,lopezperez2025multilayer}.

  \item \textbf{Extending single-array-assumed techniques to an architecture-aware
  distributed form}: CSI-acquisition, beam-management, and MIMO techniques that assumed a
  co-located array must operate across heterogeneous hardware. Conventional fully digital
  hardware retains flexibility, whereas DMA- or FAS-based tri-hybrid hardware adds
  antenna/EM processing where a large aperture is useful with few RF
  paths~\cite{triHybridMimo,dmaArray}.

  \item \textbf{Efficient hierarchical processing and effective channel construction}:
  channel acquisition, precoding, and beamforming should follow
  RU-activation sparsity and operate on effective dimensions after local antenna/EM and
  analog processing, rather than transport all element-level channels. Converter
  resolution, local processing, and fronthaul should be matched so power does not erase
  RF-chain savings~\cite{lowResAdc,bjornsonCellFree}.

  \item \textbf{A distributed structure that accommodates variable coverage and sensing}:
  on top of this architecture-aware distributed structure, common and UE-specific
  coverage, narrow communication beams, and wide-region or multiview sensing should be
  realized by selecting the participating RUs and their transmit or receive roles. This
  adds timing, phase, and location calibration across RUs and dynamic reconfiguration of
  the cooperative beam~\cite{bjornson2025gmimo,Masouros2022beyond6G}.
\end{itemize}

At a scale of 256 or 512 ports, the CSI must be more refined. The cost of acquiring it,
namely the reference-signal and feedback overhead, nevertheless grows sharply. The last category therefore asks for more accurate CSI at lower overhead.
\begin{itemize}
  \item \textbf{A design that reduces the effective dimension together with the
  reference-signal and report resources}: at a large number of antenna ports, the downlink
  reference signals and the feedback resources grow in proportion. A technique is
  therefore needed that groups the users and reduces the effective dimension through
  common beamforming, which cuts the downlink reference-signal and the report resources
  together~\cite{adhikary2013jsdm}.

  \item \textbf{Acquisition that overcomes low SNR, channel variation, and the
  downlink--uplink split}: the beamforming gain can be exploited only once the CSI is
  available, but at the estimation stage that gain is absent. The link therefore takes the
  increased path loss and the reduced per-element SNR as they are. The channel also
  changes between the time the CSI is reported and the time it is used for scheduling, so
  the CSI grows stale. Overcoming this low SNR and this channel variation calls for
  integrating the downlink and uplink through their angle--delay partial reciprocity and
  through channel prediction. The codebook mismatch from near-field spherical waves and
  spatial non-stationarity is treated as a secondary concern~\cite{partialReciprocity,nfsns2025}.

  \item \textbf{Overcoming CSI uncertainty from channel-environment information for {PMI},
  link adaptation, and scheduling}: on the refinement side, a staged approach is needed.
  First, a more accurate PMI is obtained with AI/ML. Second, the MCS-determining CQI is
  matched to the actual channel environment through outer-loop link adaptation. Third, the
  channel-environment (distribution) information is learned more broadly, so as to meet a
  target outage even under the remaining CSI uncertainty in scheduling and resource
  allocation~\cite{csiNet,aiEmimoSched}.


  \item \textbf{Acquiring long-term channel statistics}: slowly varying spatial
  channel structure can serve as reusable side information across CSI occasions,
  but it must be obtained before explicit channel sounding. A promising direction
  is to learn a channel scene representation from prior radio observations and
  render the long-term statistics at a target location, thereby reducing the
  overhead for subsequent CSI acquisition.
\end{itemize}

\begin{figure*}[t]
  \centering
  \subfloat[SSB across environments]{%
    \includegraphics[width=0.48\textwidth]{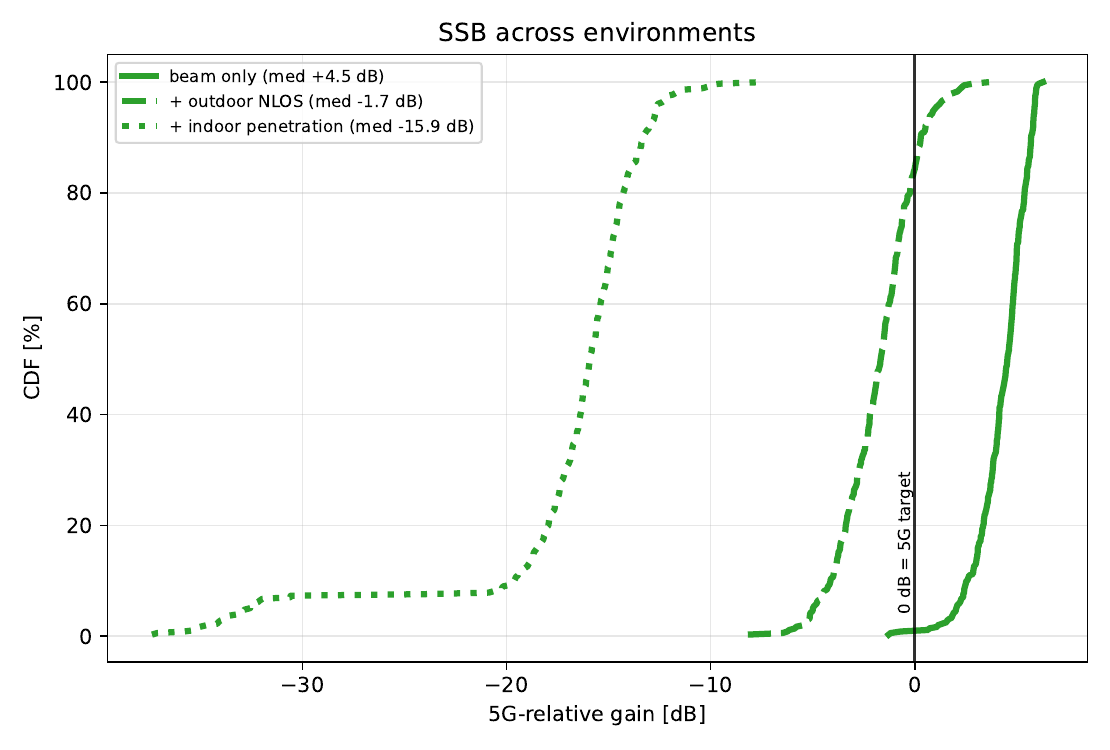}%
    \label{fig:rsrp_cdf_a}}
  \hfil
  \subfloat[Channels under the worst environment]{%
    \includegraphics[width=0.48\textwidth]{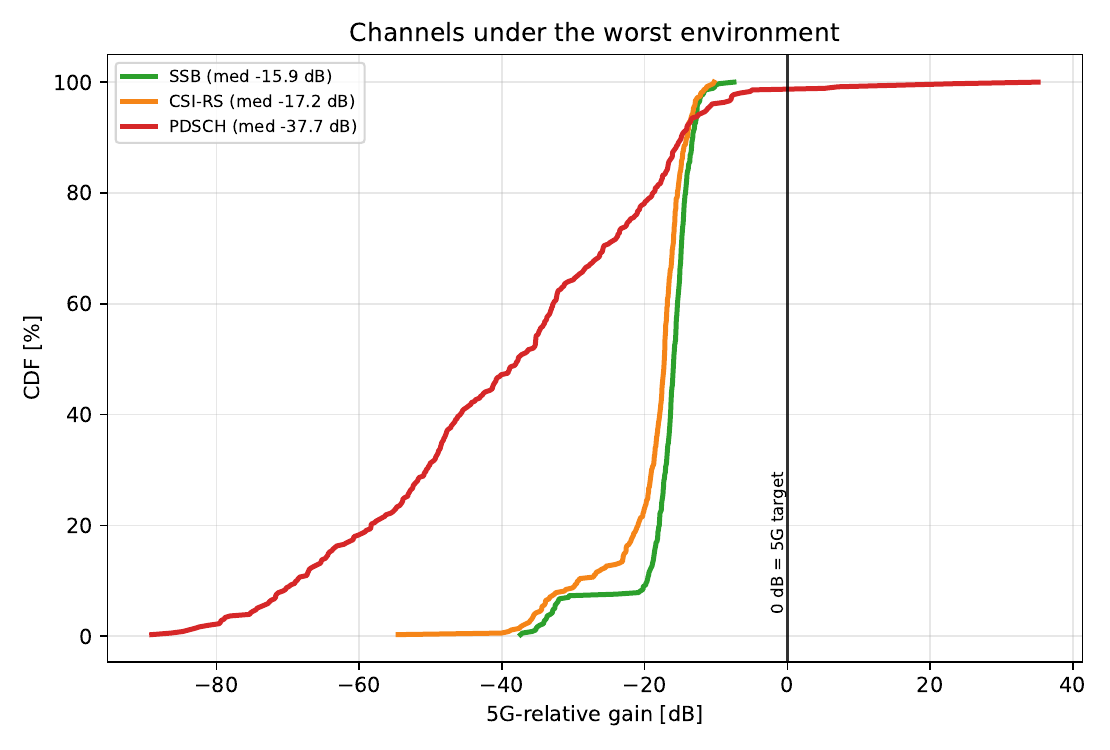}%
    \label{fig:rsrp_cdf_b}}
  \caption{CDF of the 5G-relative gain at randomly chosen user locations over the
  Herald Square ray-tracing environment of~\cite{kang2024fr3}, comparing 5G at
  3.5\,GHz with 6G at 7\,GHz using an equal-aperture antenna array at a
  co-located RU. The outdoor NLOS loss is taken from ray tracing, and the indoor
  O2I and indoor loss are generated from the 3GPP TR~38.901 Release~19 model.
  (a)~SSB, with the beam gain alone, then with the outdoor NLOS loss added, then
  with the indoor penetration loss added as well. (b)~SSB, CSI-RS, and PDSCH
  under the worst environment that combines the deep NLOS and the O2I
  penetration.}
  \label{fig:rsrp_cdf}
\end{figure*}

\section{Securing Effective Coverage Throughout the Whole Protocol}


Doubling the carrier from 3.5 to 7\,GHz raises the free-space path loss. An equal aperture compensates it, as explained above, so what remains is the excess loss
beyond free space, set not by the distance to the cell edge but by the complexity
of the environment~\cite{bjornson2025gmimo,equalAperture2026}.
Outdoors, this excess
grows as the NLOS path bends around more diffraction edges, as the blockage
deepens, and as the scattering becomes denser, reaching up to about 10 to
15\,dB in deep NLOS since all three losses rise with
frequency~\cite{shakya2024outdoor}. Indoors, a low-loss building with clear
glass and light partitions adds only a few decibels, whereas a high-loss modern
high-rise with a low-emissivity glass facade and reinforced-concrete cores
attenuates far more and scales with the building to reach up to several tens of
decibels for a deep indoor user~\cite{shakya2024penetration,o2iModern,poddar2025rel19}.
Since this environment-dependent excess is left uncompensated and is widely
distributed across the cell, securing effective coverage at an arbitrary
location is both necessary and difficult, and an equal aperture alone cannot
achieve it. 
This is the first scaling mismatch in Table~\ref{tab:mimo_scaling_breakpoints}: increasing the number of radiating elements does not automatically provide equivalent gain to every physical channel and protocol state.

The equal-aperture argument is defined on the PDSCH, which forms a UE-specific
narrow beam from the reported CSI and therefore uses the aperture at its
theoretical maximum gain~\cite{equalAperture2026,bjornson2025gmimo}. The protocol
needs several other channels for the UE in different states, and each of them
uses the aperture only in part. 
For example, the SSB is the always-on cell-common signal for
cell search and initial access in the idle state, and it covers the whole cell
with wide beams~\cite{nr38214}. The CSI-RS lets a connected UE
measure the channel and report the CSI, and it is typically not optimized per UE
but divides the cell into regions, so it uses a directional beam that splits each
SSB direction into several narrower beams~\cite{nr38214}. The SRS carries the
uplink sounding, and although its receive beamforming could in principle reach
the same maximum, the base station has no channel information yet, so it can only
use a directional beam at the CSI-RS level~\cite{bjornson2025gmimo}. Even the
PDSCH and the PUSCH, which target the theoretical narrow-beam gain, often fail to
use it in full because of the imperfection of the acquired CSI and the use of
MU-MIMO~\cite{partialReciprocity,bjornson2025gmimo}. Each of these channels
participates in forming the effective cell coverage in its own way.

Considering the equal-aperture coverage limit and the per-channel beam-gain usage
together changes the outlook. Under the physical assumption of the maximum
aperture gain, which is the PDSCH best case, the environment-dependent excess
loss makes the coverage target hard to meet over the whole
cell~\cite{equalAperture2026}. The per-channel usage above, however, shows that
most channels do not yet use that aperture, and by exploiting both the
environment and the protocol operation of the 5G physical channels there is far
more room to meet the target than the full-aperture view alone suggests. The
indoor case still cannot be met everywhere, but with more active beam management
and protocol-aware operation of these under-used channels, the coverage can be
secured without greatly increasing the need for a distributed antenna system
relative to 5G~\cite{shakya2024indoor}.

\begin{table*}[t]
\centering
\caption{Simulation scenario used to obtain the results in Fig.~\ref{fig:rsrp_cdf}.}
\label{tab:fig2_simulation_scenario}
\renewcommand{\arraystretch}{1.18}
\setlength{\tabcolsep}{5pt}

\begin{tabularx}{\textwidth}{
    >{\raggedright\arraybackslash}p{0.19\textwidth}
    >{\raggedright\arraybackslash}X}
\toprule
\textbf{Category} & \textbf{Simulation setting} \\
\midrule

\textbf{Deployment environment}
&
Herald Square terrain and three-dimensional building model, following
the ray-tracing methodology in~\cite{kang2024fr3}.
The model includes site-specific building materials and foliage.
\\

\addlinespace
\textbf{Base-station deployment}
&
A total of 18 rooftop base stations are deployed with an inter-site
distance of approximately 120~m. Each site consists of three sectors,
and the antenna downtilt is set to $-12^{\circ}$.
The per-element radiation pattern follows 3GPP TR~37.840.
\\

\addlinespace
\textbf{Carrier frequencies and arrays}
&
The 5G reference system operates at 3.5~GHz, whereas the 6G system
operates at 7~GHz. Uniform rectangular arrays are scaled with frequency
while maintaining the same physical aperture. Consequently, the 7-GHz
array contains approximately four times as many antenna elements as the
3.5-GHz array and provides approximately 6~dB higher maximum array gain.
\\

\addlinespace
\textbf{User distribution}
&
Outdoor and indoor users are randomly distributed with a ratio of
$1:1$. The serving RU is co-located for the 3.5-GHz and 7-GHz systems
to enable a location-by-location comparison.
\\

\addlinespace
\textbf{Outdoor propagation}
&
The site-specific outdoor path loss and additional NLOS loss are
obtained directly from Wireless InSite ray tracing over the Herald
Square 3D model.
\\

\addlinespace
\textbf{Indoor propagation}
&
For an indoor user, the outdoor path loss to the building exterior is
obtained from ray tracing. Outdoor-to-indoor penetration loss and indoor
loss are subsequently generated using the 3GPP TR~38.901 Release~19
model. Low-loss buildings use clear glass and concrete, whereas
high-loss buildings use infrared-reflective glass and concrete.
\\

\addlinespace
\textbf{Indoor-loss model}
&
The indoor loss is modeled as
\[
    PL_{\mathrm{in}} = 0.5 d_{\mathrm{2D\mbox{-}in}},
\]
where $d_{\mathrm{2D\mbox{-}in}}$ is determined from the building size.
A log-normal penetration term is added with standard deviation
$4.4$~dB for low-loss buildings and $6.5$~dB for high-loss buildings.
\\

\addlinespace
\textbf{5G beam configuration}
&
At 3.5~GHz, the SSB is transmitted without beamforming. The CSI-RS
partitions the cell using directional beams with gains of up to 6~dB.
\\

\addlinespace
\textbf{6G SSB configuration}
&
At 7~GHz, the SSB covers the cell using spatially partitioned beams
with gains of up to 6~dB over the equal physical aperture. Its beam
configuration therefore corresponds approximately to that of the
3.5-GHz CSI-RS.
\\

\addlinespace
\textbf{6G CSI-RS configuration}
&
Each 7-GHz SSB-beam region is further subdivided using CSI-RS beams
that provide up to an additional 6~dB gain, resulting in a maximum
beam gain of approximately 12~dB. The subdivision is intentionally
kept coarse to limit CSI-RS resource overhead.
\\

\addlinespace
\textbf{PDSCH transmission}
&
The PDSCH employs multi-user MIMO block-diagonalization precoding
for randomly scheduled users. The resulting gain accounts for
multi-user beamforming-gain sharing and imperfect CSI.
\\

\addlinespace
\textbf{Evaluation metric}
&
The performance metric is the 7-GHz gain relative to the corresponding
3.5-GHz link at the same user location:
\[
    G_{\mathrm{rel},c}
    =
    P_{\mathrm{rx},c}^{\mathrm{6G},\,7\,\mathrm{GHz}}
    -
    P_{\mathrm{rx},c}^{\mathrm{5G},\,3.5\,\mathrm{GHz}},
\]
where $c\in\{\mathrm{SSB},\mathrm{CSI\mbox{-}RS},
\mathrm{PDSCH}\}$. Here, $(P_{\mathrm{rx},c}^{\mathrm{6G}})$ and
$(P_{\mathrm{rx},c}^{\mathrm{5G}})$ denote the received powers
in decibels for physical channel $(c)$ in the 7-GHz and
3.5-GHz systems, respectively. Thus,
\(G_{\mathrm{rel},c}\) is the 5G-relative received-power gain
in decibels at the same user location. A value of $0$~dB indicates that the 6G link
meets the corresponding 5G coverage target.


\\

\bottomrule
\end{tabularx}
\end{table*}


Fig.~\ref{fig:rsrp_cdf} evaluates the 7-GHz coverage relative to the
corresponding 3.5-GHz link at the same user location, where
0\,dB indicates that the 7-GHz system meets the 5G coverage
target. Table~\ref{tab:fig2_simulation_scenario} summarizes the
simulation scenario. The comparison keeps the deployment
geometry and physical array aperture fixed across the two
carrier frequencies, thereby isolating the effects of
frequency-dependent propagation and physical-channel-specific
beamforming. Fig.~\ref{fig:rsrp_cdf}(a) progressively introduces the outdoor
NLOS and indoor penetration losses into the SSB comparison,
whereas Fig.~\ref{fig:rsrp_cdf}(b) compares the SSB, CSI-RS, and PDSCH under
the combined deep-NLOS and O2I condition.

Fig.~\ref{fig:rsrp_cdf_a} shows that the environment progressively erodes the nominal aperture-scaling gain. With beamforming alone, the 7-GHz SSB meets the 5G target at most locations, with a median relative gain of approximately 4.5\,dB. Outdoor NLOS shifts the median below the target to approximately $-1.7$\,dB, whereas the additional O2I and indoor losses produce a much heavier lower tail and a median of approximately $-15.9$\,dB. Thus, equal-aperture scaling does not compensate the environment-dependent excess loss uniformly across the cell.


Fig.~\ref{fig:rsrp_cdf_b} further shows that the coverage deficit depends strongly on the physical channel. Under the combined deep-NLOS and O2I condition, the SSB and CSI-RS exhibit median relative gains of approximately $-15.9$ and $-17.2$\,dB, respectively. The PDSCH performs substantially worse under the assumed random scheduling, block-diagonalization precoding, and imperfect CSI, with a median of approximately $-37.7$\,dB. This separation demonstrates that equal-aperture coverage cannot be assessed from a single idealized beamforming gain; it must be evaluated jointly with the beam configuration, channel-acquisition quality, and protocol operation.


We note that this study is intended to expose the coverage asymmetry rather than to represent an optimized 6G deployment. It does not optimize the CSI-RS resources, user scheduling, or PDSCH precoding, nor does it include dedicated in-building solutions for severe penetration loss. The absolute CDF values should therefore be interpreted cautiously. Nevertheless, the results show that environment-dependent loss and physical-channel-specific beamforming must be considered jointly; equal-aperture scaling alone does not guarantee effective coverage across the entire protocol.


Providing this location- and channel-specific beam gain becomes easier with
site-specific side information, such as a high-precision map and a digital twin
that reflects the measured radio environment~\cite{siteSpecific2026,freqAdaptiveMB}. Building a high-precision map
and acquiring radio information that matches the actual environment, however, are
costly, and a non-civilian dedicated network for special use changes its coverage
demand with time and place~\cite{publicSafetyMC}. Securing such coverage
therefore needs an additional capability that reconfigures the required
information automatically and quickly within the functions of the
RAN~\cite{lopezperez2025multilayer}. Sensing adds a further dimension, because
depending on whether it is monostatic or bistatic, ISAC provides a kind of
coverage that differs from that of communication~\cite{Masouros2022beyond6G}.
Securing the effective coverage that includes this sensing requires technical
progress across the several aspects of E-MIMO discussed above, namely the
efficient environment-aware coverage of the whole cell, the per-channel
equivalence of the protocol, the restoration of the indoor coverage, and the
fast and adaptive reconfiguration of variable coverage.

\section{6G E-MIMO RF Devices and RU}

To seamlessly deploy 6G E-MIMO base stations at existing 5G macro sites while maintaining the same physical aperture size and form factor, the RU architecture must adapt to the challenges of operating at FR3 bands, which are approximately twice the frequency of 5G mid-bands~\cite{cui2023fr3}. Achieving equivalent or superior coverage within a constrained spatial footprint demands not only advanced RF device technologies to overcome severe propagation and material losses but also hardware-efficient RU topologies. Consequently, substantial hardware-level research and innovations in both RF component efficiency and sub-system integration are indispensable. This section delineates the core technical challenges and critical research directions for 6G E-MIMO RF devices and RUs across four pivotal domains.

Accordingly, the radiating-elements, active-RF-chain, and
instantaneous-bandwidth rows of Table~\ref{tab:mimo_scaling_breakpoints} must be considered jointly. An architecture that reduces the number of active RF chains is useful only if its feeding-network, insertion, calibration, and wideband losses do not eliminate the expected aperture gain.


\subsection{Wideband High-Efficiency Transceiver}

Operating at the 6G FR3 band with wide bandwidths (up to 400 MHz) imposes severe hardware constraints. Unlike the mature 5G mid-band or the actively developed mm-wave beamforming IC ecosystems, circuit design for this upper mid-band is still in its infancy. RF front-end circuits, including PAs, low-noise amplifiers (LNAs), phase shifters combined with variable gain/delay circuits, frequency up/down converters and synthesizers must maintain their gain, noise, linearity, and efficiency over a wider operating band and remains a significant challenge.~\cite{Tataria2021}

The primary bottleneck of FR3 E-MIMO arises not only from the drastic scaling of antenna elements and TRX chains, but also from new 6G features such as ISAC, AI-assisted radio control, and full duplex, which add further hardware and computational overhead. To maintain the total Radio Unit (RU) power consumption within a reasonable budget, individual device-level efficiency is critical. Specifically, per-antenna PAs, per-TRX ADCs/DACs and RF chain components must be optimized for ultra-low power. Without highly efficient individual devices, the cumulative power of massive RF chains would scale unsustainably, making semiconductor device-level energy innovation an essential research direction.

\subsection{Multi-Beam Network Based RU Architecture}

\begin{figure*}[t]
    \centering
    \subfloat[]
    {
        \includegraphics[height=0.260\textwidth]{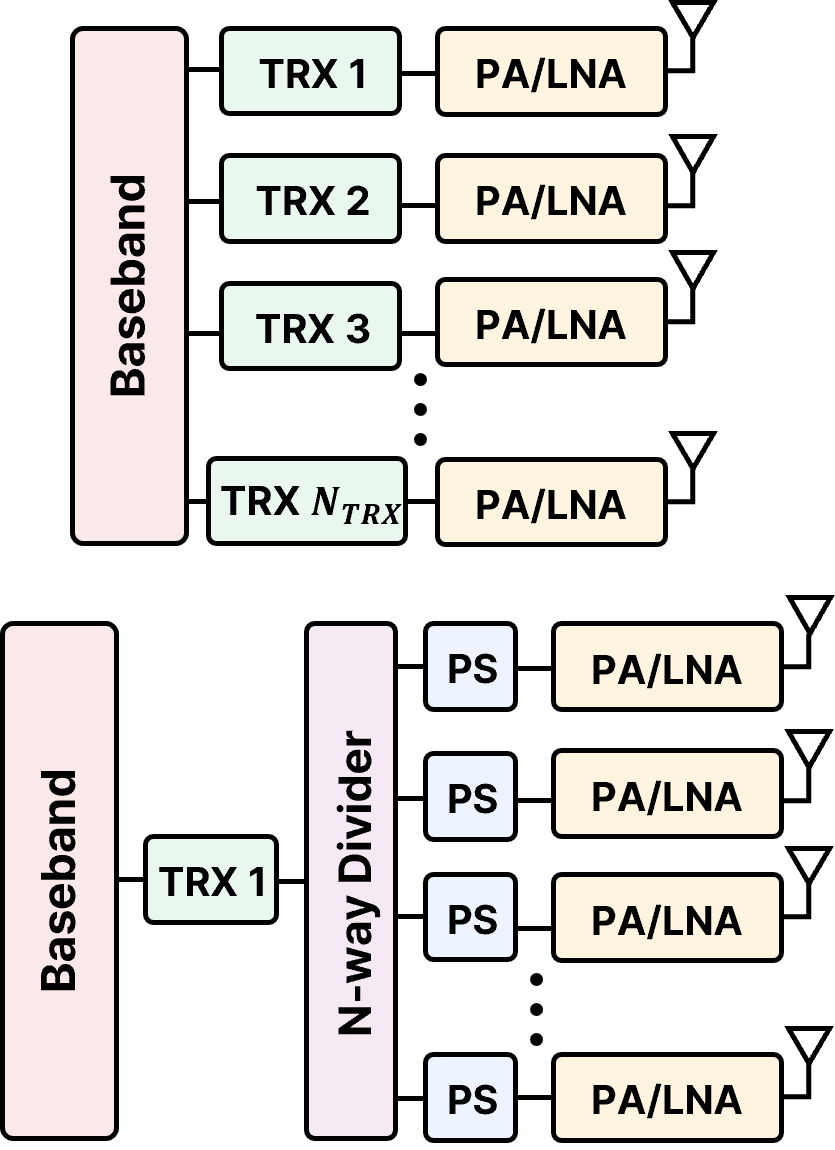}
        \label{struct_a}
    }
    \hfil
    \subfloat[]
    {
        \includegraphics[height=0.250\textwidth]{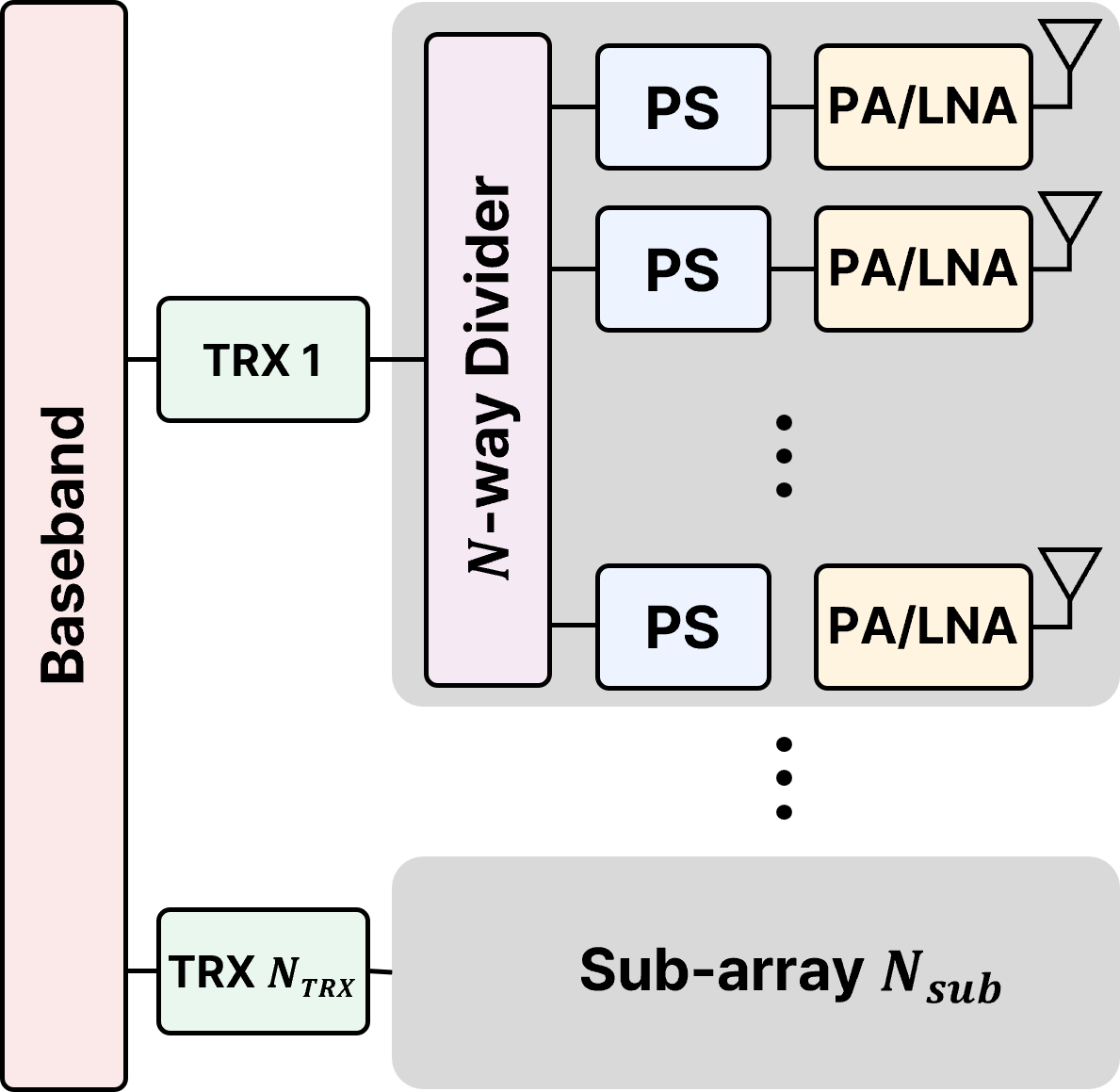}
        \label{struct_c}
    }
    \hfil
    \subfloat[]
    {
        \includegraphics[height=0.250\textwidth]{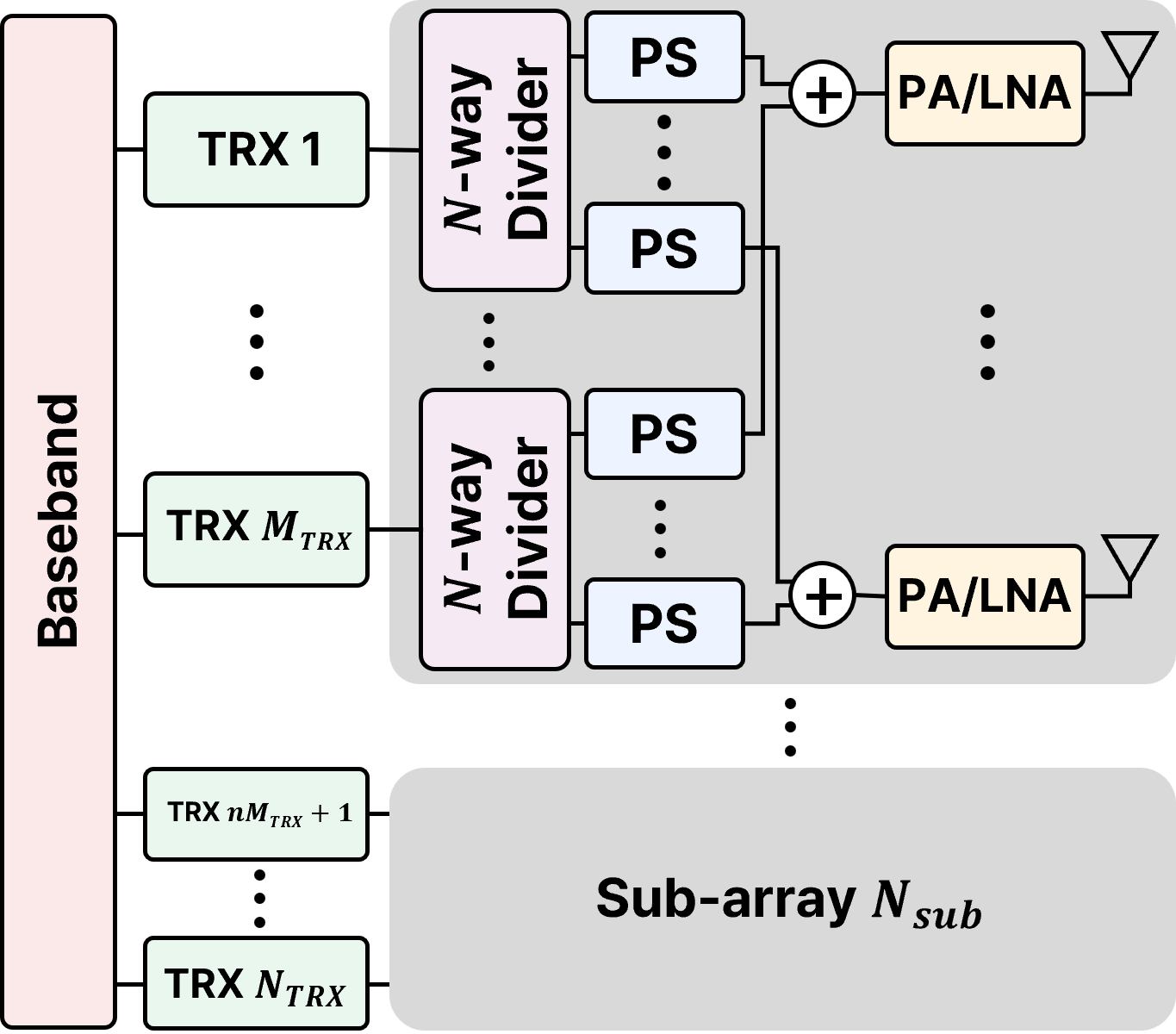}
        \label{struct_d}
    }
    \par\vspace{2mm}
    \subfloat[]
    {
        \includegraphics[height=0.250\textwidth]{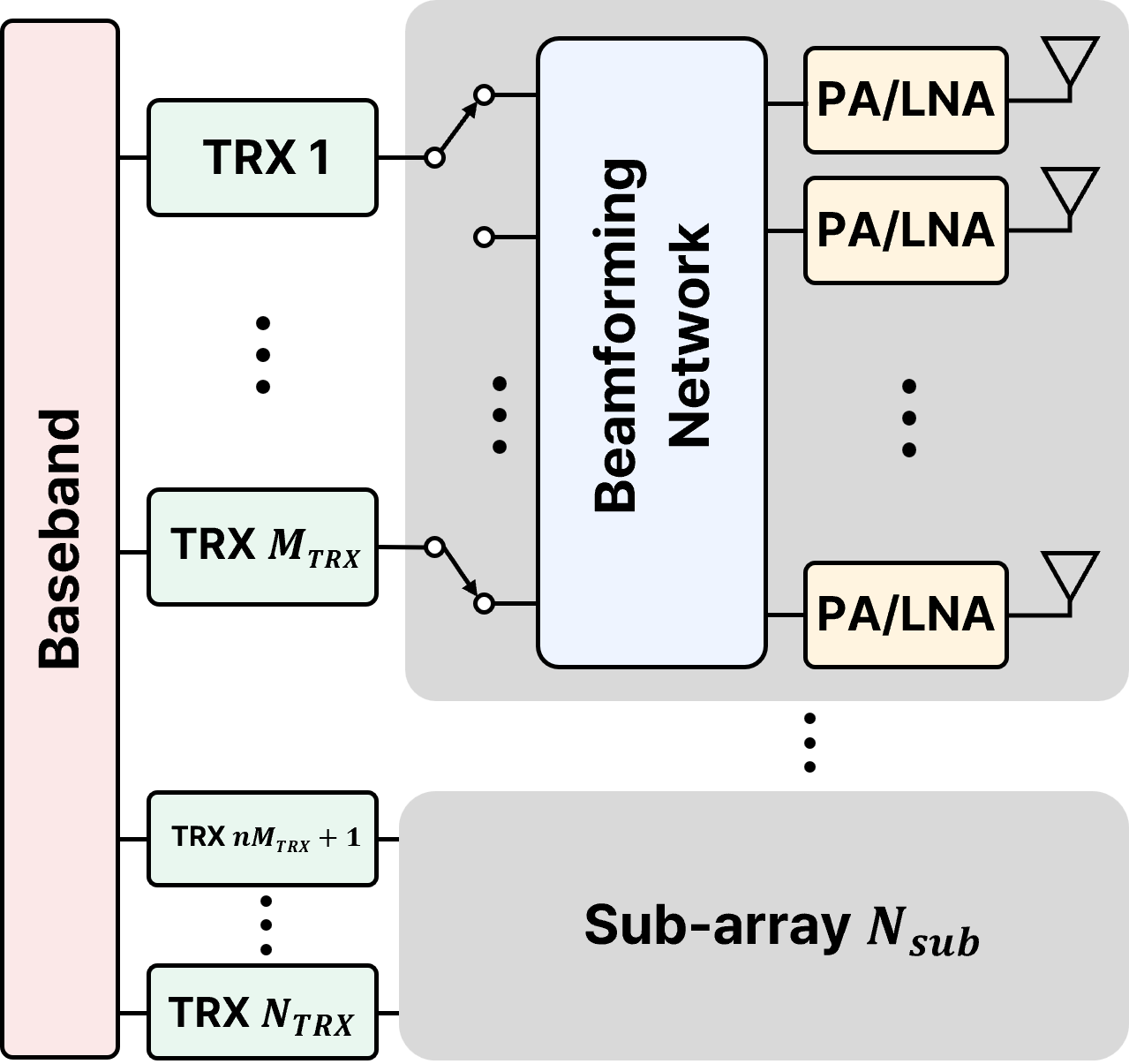}
        \label{struct_e}
    }
    \hfil
    \subfloat[]
    {
        \includegraphics[height=0.250\textwidth]{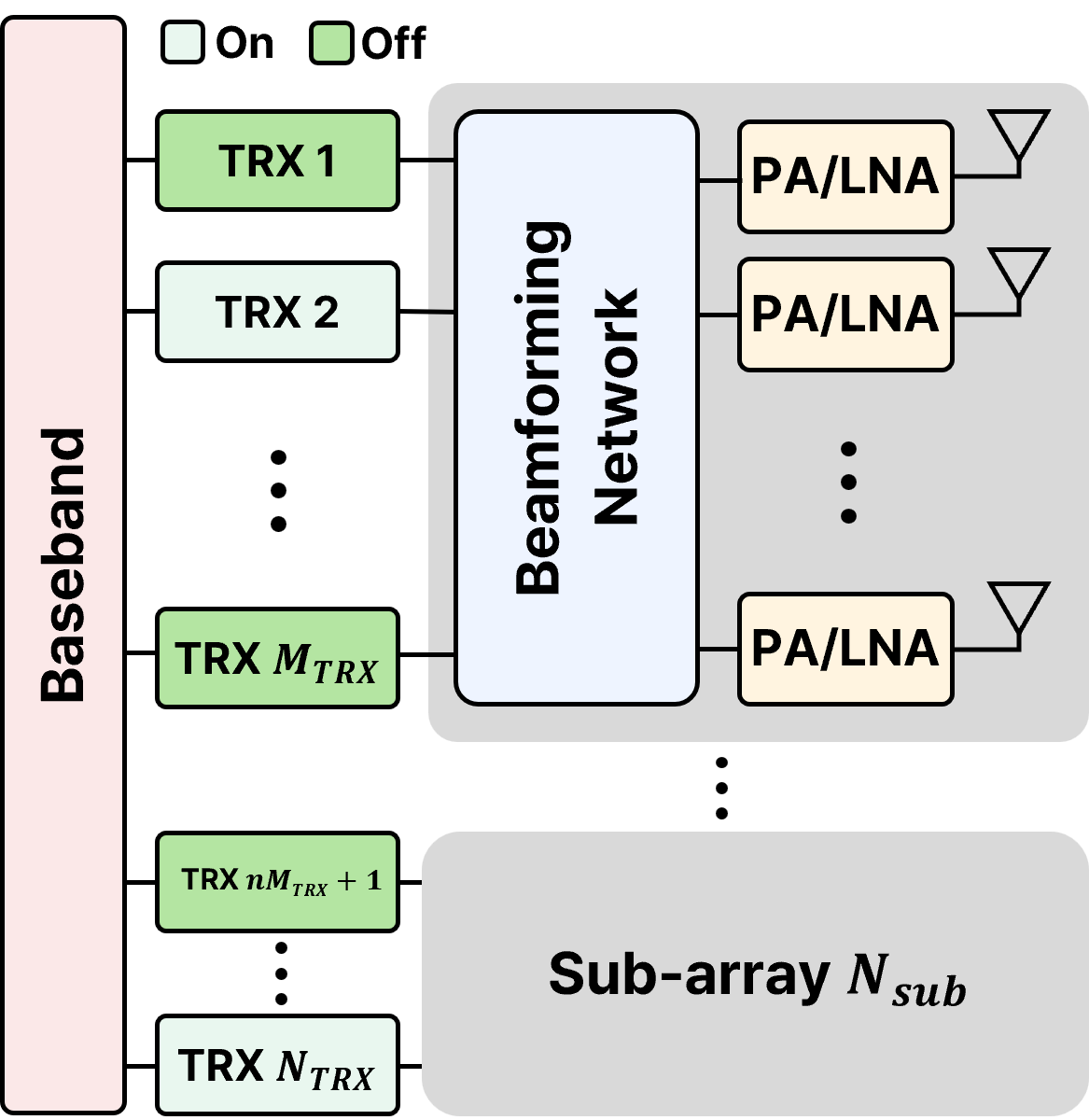}
        \label{struct_f}
    }
    \hfil
    \subfloat[]
    {
        \includegraphics[height=0.250\textwidth]{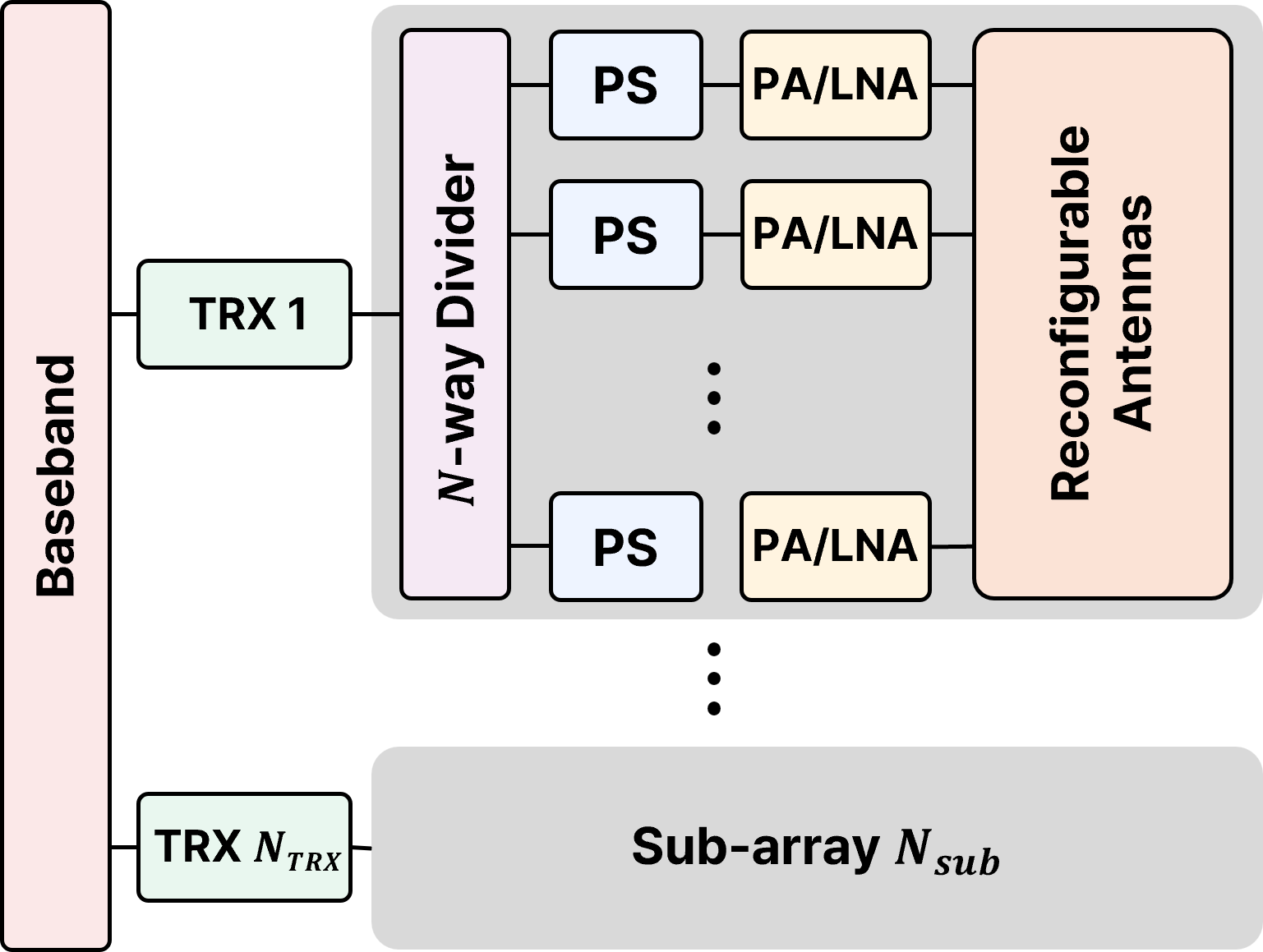}
        \label{struct_g}
    }
    \caption{Progression of beamforming architectures from fully digital beamforming and fully analog beamforming to hybrid beamforming structures: (a) fully digital beamforming (above) and fully analog beamforming (below), (b) phase-shifter-based hybrid beamforming with a single RF chain per subarray, (c) phase-shifter-based hybrid beamforming with multiple RF chains per subarray, (d) fixed-BFN-based hybrid beamforming with a relatively small number of TRXs and an input switching network, (e) fixed-BFN-based hybrid beamforming with a large number of TRXs and selective RF-chain activation, and (f) tri-hybrid beamforming with EM layer composed of reconfigurable antennas.}
    \label{fig:beamforming_evolution}
\end{figure*}

Beamforming architectures can be broadly classified into digital, analog, and hybrid approaches. Fully digital beamforming provides the highest degrees of freedom and multi-user scaling, but requires a dedicated transceiver (TRX) chain for each antenna element, leading to prohibitive power consumption and hardware costs. In contrast, analog beamforming uses a single transceiver chain connected to multiple antennas via active phase shifters, which significantly reduces cost but offers limited multi-beam flexibility. To balance these trade-offs, 5G mid-band massive MIMO systems have widely adopted hybrid beamforming structures, combining a digital beamforming with subarray-level analog beamforming based on fixed analog subarray beams to achieve high spectral efficiency with manageable hardware complexity~\cite{Ahmed20185G}.

However, in 6G Extreme-MIMO (E-MIMO) systems, the number of antenna elements may scale to more than one thousand, while the number of transceiver (TRX) chains must remain limited to stay within a reasonable power consumption and thermal budget. This architecture naturally increases the number of antenna elements assigned per subarray, which in turn sharpens the beamwidth of the subarray. Consequently, dynamic beam-steering capabilities must be integrated into the subarray level. Furthermore, to support multiple spatially distributed users simultaneously, multiple TRX chains must be connected to a subarray which can support multiple beams. 

These evolutionary hybrid structures can be systematically analyzed through the topologies illustrated in Fig.~\ref{fig:beamforming_evolution}, which correspond to analog, 5G hybrid, and digital beamforming architectures in Fig.~\ref{fig:beamforming_evolution}(a), (b), and (f), respectively. To handle the multi-user and multi-beam requirements under tight power constraints, a multi-beam subarray configuration as shown in Fig.~\ref{fig:beamforming_evolution}(c) becomes highly desirable. In this context, achieving analog multi-beam generation conventionally demands a fully connected subarray topology. Let $N_{ant}$ and $N_{TRX}$ denote the total number of antenna elements and TRX chains in the RU, respectively, while $M_{ant}$ and $M_{TRX}$ represent the number of antennas and TRX chains allocated per subarray, satisfying $N_{\mathrm{TRX}}=N_{\mathrm{sub}}M_{\mathrm{TRX}}$. In a fully connected subarray, each of the $M_{ant}$ antennas must interface with all $M_{TRX}$ chains, which leads to an exponentially high hardware complexity and routing overhead. To overcome this practical bottleneck, hardware-efficient alternatives such as Fig.~\ref{fig:beamforming_evolution}(d) and (e) are proposed.

\begin{figure}[t]
    \centering
    \subfloat[]
    {
        \includegraphics[
            height=0.230\textwidth,
            trim=1mm 1mm 1mm 1mm,
            clip
        ]{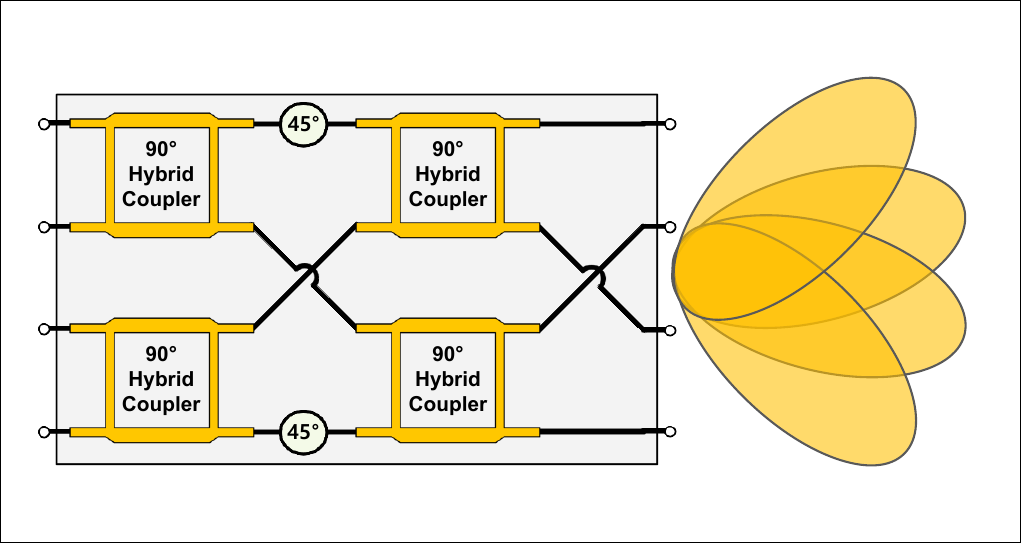}
        \label{struct_a}
    }
    \par\vspace{0mm}
    \subfloat[]
    {
        \includegraphics[
            height=0.22\textwidth,
            trim=1mm 1mm 1mm 1.5mm,
            clip
        ]{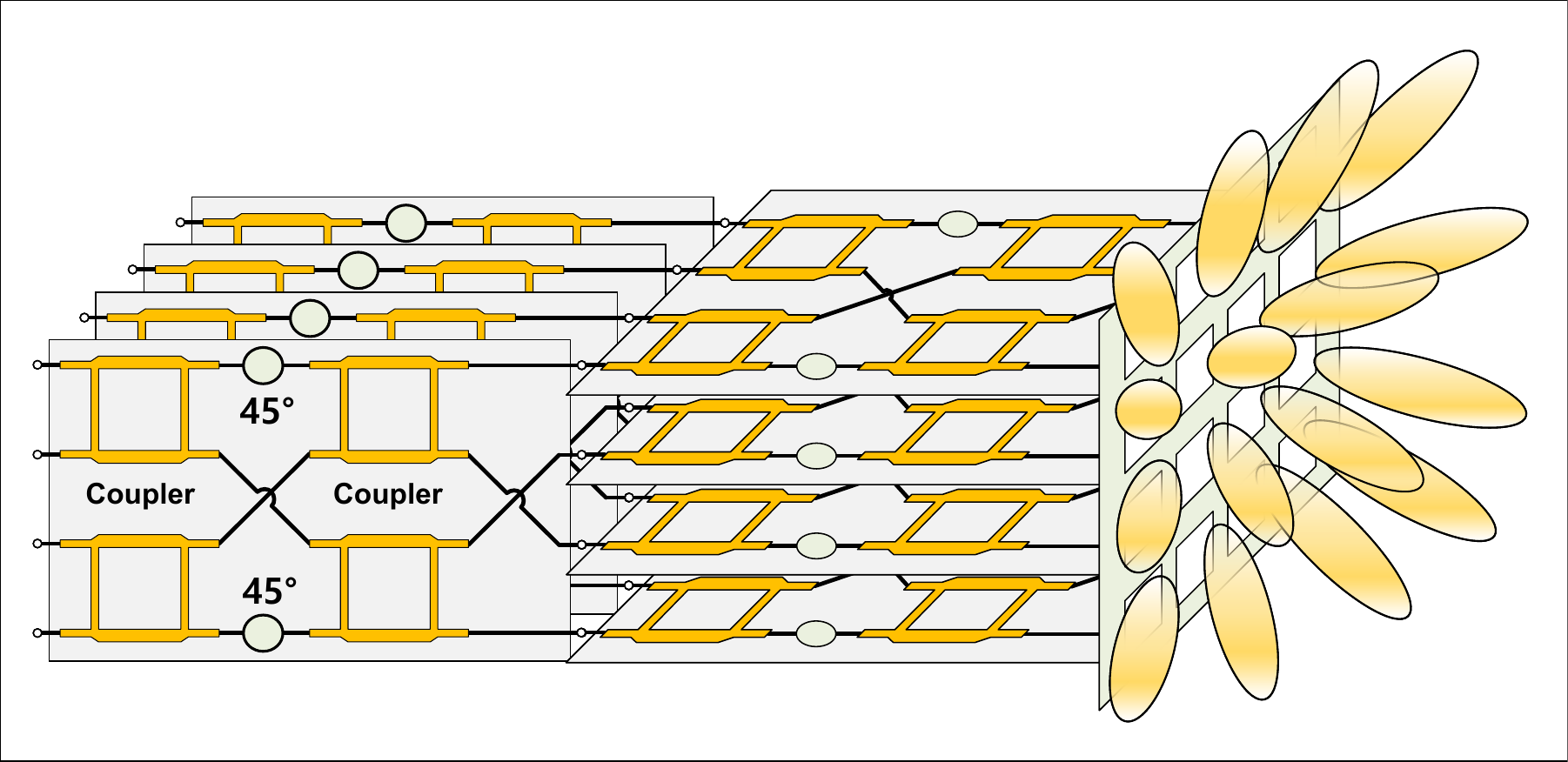}
        \label{struct_e}
    }
    \hfil
    \caption{Butler matrix for one- and two-dimensional antenna arrays: (a) a \(4\times4\) Butler matrix feeding a \(4\times1\) linear array to generate four fixed beams in one angular dimension, and (b) a two-dimensional \(16\times16\) Butler matrix, implemented by cascading two orthogonal stages of \(4\times4\) Butler matrices, feeding a \(4\times4\) planar array to generate separable beams in the azimuth and elevation dimensions.}
    \label{fig:Butler_Matrix}
\end{figure}

Advanced passive Beamforming Networks (BFNs), typified by the Butler matrix (Fig.~\ref{fig:Butler_Matrix}~\cite{Butler1961Beam}) as a representative architecture, serve as a compelling solution, with the topology in Fig.~\ref{fig:beamforming_evolution}(d) offering distinct architectural advantages over the conventional fully connected array in Fig.~\ref{fig:beamforming_evolution}(c). Fundamentally, a passive BFN pre-generates a set of mutually orthogonal fixed beams rather than dynamically synthesizing individual beams toward arbitrary directions on demand. However, by properly combining and overlapping these pre-defined orthogonal beams, this architecture can effectively synthesize a new directional beam toward any arbitrary angle, enabling continuous beam coverage~\cite{Lee2023BM}. By utilizing these passive structures like the Butler matrix, Blass matrix, Rotman lens, or dielectric lens, this configuration enables robust multi-beam synthesis without relying on active phase shifters~\cite{Chae2016lens}. This phase-shifterless beamforming drastically reduces insertion losses, simplifies control circuitry, and mitigates power-unbalance issues, while still offering excellent tracking flexibility through a switch network outside the rigid boresight direction.

Furthermore, the alternative topology presented in Fig.~\ref{fig:beamforming_evolution}(e) introduces advanced degrees of freedom through hardware scalability and computational efficiency. As $M_{TRX}$ approaches $M_{ant}$, this architecture becomes equivalent to the fully digital configuration of Fig.~\ref{fig:beamforming_evolution}(f) with an integrated passive BFN, thereby inheriting identical beamforming capabilities. Although this regime inherently increases $N_{TRX}$ and baseline power consumption, Fig.~\ref{fig:beamforming_evolution}(e) effectively mitigates this through traffic-adaptive control; depending on the spatial distribution of active users, the RU can selectively activate only a subset of TRX chains and power down unused ones via power-gating. Beyond this dynamic power management, this structure offers a profound advantage in reducing digital baseband complexity. Because the analog BFN inherently manages high-complexity phase adjustments, the digital processor is relieved from intensive complex-number matrix computations. Instead, the digital domain can optimize beamforming weights using low-complexity real-number computations focused primarily on magnitude coefficients, which substantially lowers both processing latency and baseband power dissipation for dense 6G E-MIMO deployments.

\subsection{Low-Power and High-Linearity Power Amplifiers}

Future FR3 E-MIMO radio units employ hundreds of dense power amplifiers (PAs) operating over instantaneous bandwidths exceeding $400\text{~MHz}$ while supporting high-order modulation schemes with high peak-to-average power ratio (PAPR) waveforms. Consequently, these PAs must simultaneously satisfy stringent linearity requirements---such as adjacent channel leakage ratio (ACLR) and error vector magnitude (EVM) metrics---while maintaining high energy efficiency under deep output power back-off (OBO) conditions. Because linearity and efficiency are inherently conflicting paradigms, mitigating this trade-off has become a primary bottleneck for future 6G RF front-ends.

At the FR3 band, the technical gap is particularly pronounced as systems transition between sub-6 GHz Lateral Diffused MOS (LDMOS) technologies and millimeter-wave SiGe/CMOS or GaN ecosystems. GaN-on-Si or GaN-on-SiC devices offer high power density at 7--8 GHz, yet their nonlinear electro-thermal modeling is less mature than that of legacy LDMOS. To enforce high linearity, digital predistortion (DPD) combined with crest factor reduction (CFR) remains the dominant compensation technique \cite{Wang2020DPD, Kobal2022DPD}. However, applying conventional DPD across 256 or more independent ports over a $400\text{~MHz}$ bandwidth causes a computational explosion in the digital baseband and requires unsustainable oversampling rates for the feedback loop receivers. Therefore, low-complexity, block-structured, or AI-assisted DPD architectures that exploit spatial correlation among the array paths are promising direction.

On the efficiency enhancement front, advanced architectural innovations must be tightly integrated into the RU. Envelope tracking (ET) improves efficiency by dynamically modulating the PA supply voltage according to the instantaneous signal envelope \cite{Bhardwaj2023ET, Hsu2021ET}. For broadband 6G signals, this demands wideband, high-efficiency hybrid supply modulators capable of tracking multi-hundred megahertz envelopes without introducing additional spectral regrowth \cite{Bhardwaj2023ET, Hsu2021ET}. Alternatively, outphasing and polar transmitter architectures enhance efficiency by decomposing high-PAPR waveforms into constant-envelope signals, enabling PAs to operate continuously in highly efficient saturated regions \cite{Barton2016Outphasing, Ning2020Outphasing, Kavousian2008Polar, Hu2023Polar}. Meanwhile, Doherty power amplifiers (DPAs) remain a crucial solution for active load modulation \cite{Camarchia2015Doherty, Park2022Doherty}, with current research focusing on broadband, ultra-compact topologies that can be seamlessly packed into a dense, tightly pitched phased-array form factor without causing severe mutual coupling or thermal hotspots.

\subsection{Integrated Communication-and-Sensing RF Array}
Upper-mid-band E-MIMO arrays are no longer restricted to pure communication pipelines. Under the integrated sensing and communication (ISAC) paradigm, radar-like environmental sensing and high-speed data transmission natively share the same RF hardware components and spatial degrees of freedom, yielding substantial integration and coordination gains~\cite{Masouros2022beyond6G}. However, embedding dual-functional capabilities into a highly scaled array imposes severe constraints on the RF and transceiver design, primarily governed by self-interference and coverage limitations.

The primary hardware bottleneck stems from self-interference during monostatic sensing or in-band full-duplex (IBFD) operations, where the array transmits and receives simultaneously on the same time-frequency resources. Unlike traditional full-duplex communication networks where the entire self-leakage signal is completely nullified, ISAC receivers must ensure that the weak target echo survives the cancellation process, as the echo itself constitutes the desired signal for sensing~\cite{Tang2024FDRISAC}. Achieving sufficient isolation between transmit and receive sub-arrays or channels thus requires a multi-layered cancellation strategy, including physical antenna separation, analog-domain RF cancellation circuits, digital-domain adaptive filtering, and spatial null-steering beamforming~\cite{MIN2022ISACSIC, MIN2026FDR}. The efficacy of this multi-stage cancellation sets the fundamental limit on the minimum detectable radar cross-section (RCS) in multi-panel and hybrid sub-array topologies.

The second major challenge relates to spatial coverage and geometric shadowing. Since macro base stations radiate downward from elevated positions, physical blockages in complex urban or indoor environments create severe sensing and communication blind spots. To overcome these non-line-of-sight (NLOS) limitations, Reconfigurable Intelligent Surfaces (RIS) can be dynamically deployed to reshape the propagation environment, providing a virtual line-of-sight (LOS) path that extends both data coverage and the radar sensing field-of-view~\cite{Liu2023RIS}. Hardware-efficient implementations for the upper-mid band have already demonstrated this feasibility; for instance, a scalable 1-bit RIS architecture operating at the $8\text{~GHz}$ band has been realized with low insertion loss and independent dual-polarization control~\cite{Kim20246GRIS}. Integrating such environment-aware surfaces with the E-MIMO radio unit allows the system to seamlessly toggle between narrow communication beams and wide-region radar sweeps, unlocking the full potential of Day-1 ISAC capability in 6G networks.

\section{6G E-MIMO Low-Power Consumption}


Improving device and RU efficiency can contain the power consumed within an E-MIMO panel. Such component-level advances alone, however, do not remove the power required to overcome a long or deeply blocked RU-to-UE link. In a co-located array, all radiating elements experience nearly the same large-scale attenuation. Increasing the number of elements can provide coherent array gain, but it does not shorten the propagation distance or create an alternative path around a blockage. Distributed MIMO (DMIMO) changes this geometry by partitioning the installed aperture among spatially separated RUs. A user is then likely to have a favorable large-scale channel to at least one RU, so proximity and macrodiversity can reduce the radiated power needed to meet a target link quality. This benefit does not make an indiscriminately dense distributed deployment energy efficient. The static power of additional RUs and the costs of fronthaul, synchronization, calibration, and distributed processing can erase the propagation gain. Low-power DMIMO therefore requires joint control of where the aperture is placed, which RUs are active, how each active RU realizes its local aperture, and how the resulting channels are processed. This section organizes these requirements into four research directions.

\subsection{Distributed Apertures for Propagation-Aware Power Reduction}

The aperture-distribution row of
Table~\ref{tab:mimo_scaling_breakpoints} highlights that DMIMO does not remove the scaling cost but transfers part of it from propagation and PA output power to infrastructure, fronthaul, synchronization, and calibration. The primary low-power advantage of DMIMO originates from the large-scale channel rather than from a new beamforming law. A co-located array can increase its local aperture gain, but all of its elements remain subject to nearly the same site-dependent distance, penetration loss, and dominant blockage. A distributed aperture instead creates geometrically distinct RU-to-UE links. For a given user, a nearby or less obstructed RU can therefore reduce the radiated power required to satisfy a common coverage and quality-of-service target. DMIMO should be understood as a mechanism for converting installed spatial diversity into proximity and macrodiversity gains, not as a requirement that all RUs coherently serve every UE~\cite{ngoCellFree2017}.

RU locations should be planned under this power objective rather than by uniform densification. A candidate site is valuable when it creates a favorable link for users that otherwise require high radiated power because of penetration loss, persistent blockage, or concentrated traffic. Its value is smaller when it only duplicates already strong links. Radio-environment and traffic information can therefore be used to compare the expected propagation saving with the RU's standby, activation, and transport costs. The relevant metric is the end-to-end energy required to meet a common quality-of-service target, rather than transmit power under unequal coverage or rate~\cite{ngoEnergyEfficiency2018}.

The propagation saving translates into lower network power only when the active infrastructure is kept sparse. Each powered RU incurs a baseline cost before radiating useful energy, and this cost becomes important as the number of candidate sites grows~\cite{ngoEnergyEfficiency2018}. A scalable architecture should deploy enough candidate RUs to offer favorable geometry, but activate the smallest user-centric subset that satisfies the coverage, rate, reliability, and latency targets~\cite{buzziUserCentric2020}. RU association is consequently coupled to sleep control and power allocation. Candidate clusters may be selected from slowly varying large-scale channel information and traffic statistics, after which only the necessary RUs and active paths are scheduled on the faster time scale.

The cooperation mode must be selected with the same power accounting. Coherent joint transmission can increase array and interference-management gains, but it requires inter-RU phase alignment, reciprocity calibration, and low-latency information exchange. When the incremental beamforming gain is smaller than this coordination cost, RU selection, noncoherent transmission, or partially coherent cooperation may be preferable. Cell-free and user-centric processing provide useful principles for limiting each UE and RU to a bounded neighborhood instead of forming one network-wide array~\cite{ngoCellFree2017,buzziUserCentric2020}. The central question is thus not how to keep the largest distributed array active, but how to expose many candidate apertures while energizing only the spatial resources that materially improve the link.

\subsection{Architecture-Aware Operation of Heterogeneous RUs}

\begin{figure}[t]
    \centering
    \subfloat[]
    {
        \includegraphics[width=0.48\textwidth]{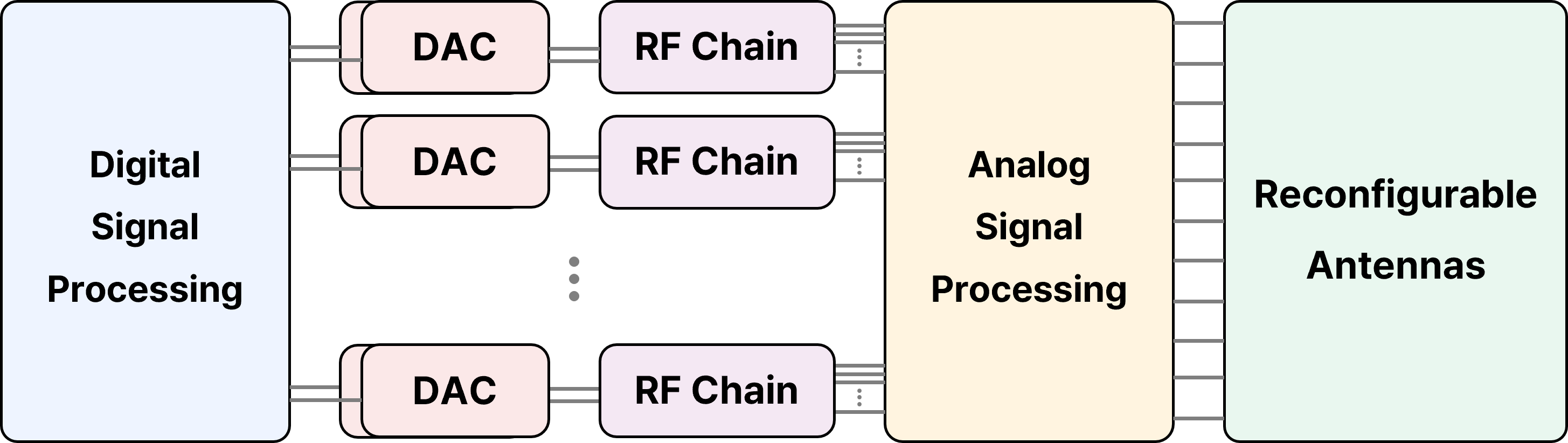}
        \label{tri_arch1}
    }
    \hfil
    \subfloat[]
    {
        \includegraphics[width=0.48\textwidth]{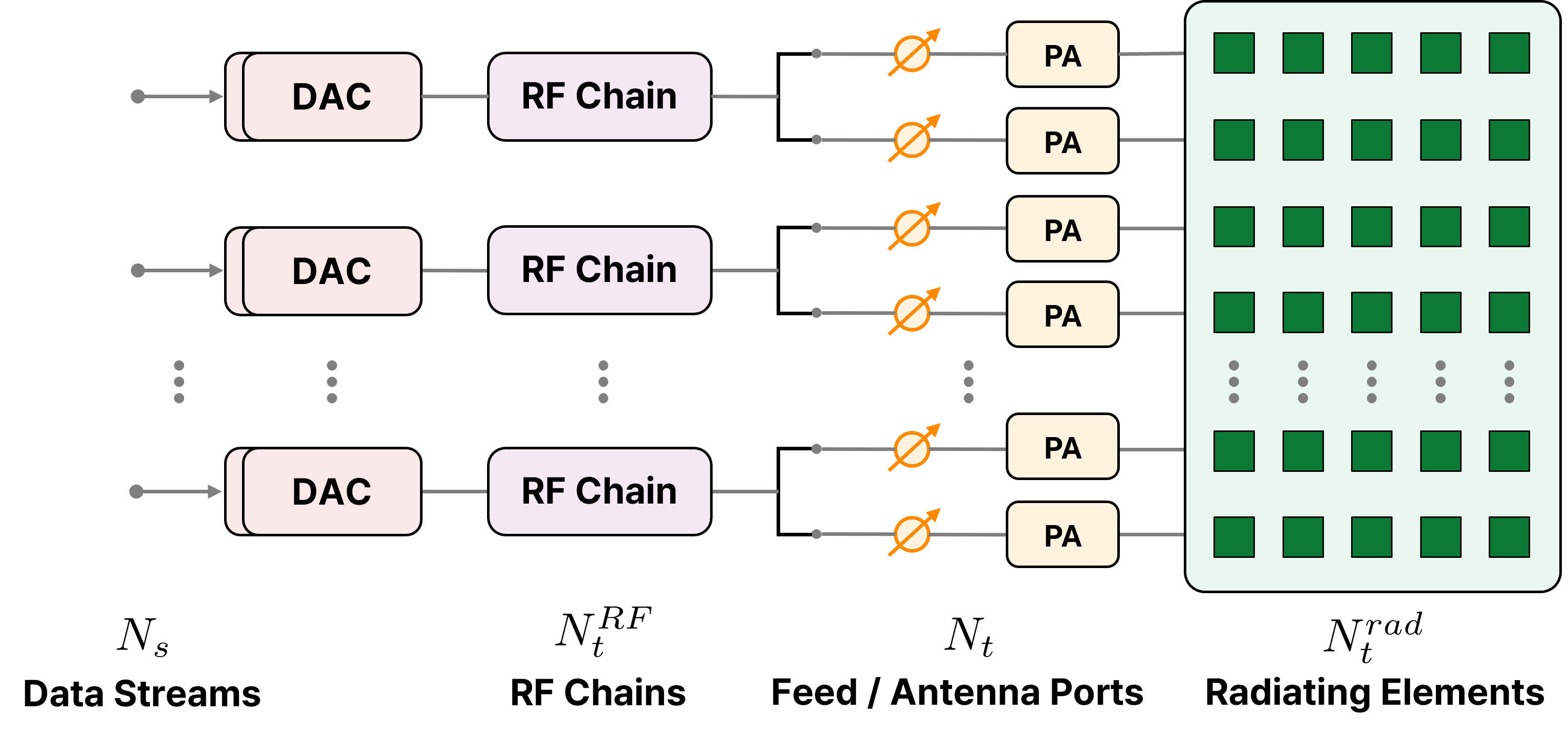}
        \label{tri_arch2}
    }
    \caption{General tri-hybrid E-MIMO architecture. (a) The transmitter combines digital precoding, analog precoding, and antenna domain or EM domain reconfiguration. The reconfigurable aperture can be realized using different hardware structures, including dynamic metasurface antennas, reconfigurable metasurfaces, parasitic arrays, fluid antenna arrays, or other aperture-reconfigurable implementations. (b) Hardware-level view of the corresponding signal path, where the \(N_{s,m}\) data streams are mapped to
\(N_{\mathrm{RF},m}\) active RF chains, processed through
\(N_{\mathrm{port},m}\) antenna/feed ports, and coupled to
\(N_{\mathrm{rad},m}\) physical radiating elements by the
reconfigurable aperture.}
    \label{fig:tri_hybrid_architecture}
\end{figure}

DMIMO specifies the network geometry, but it does not prescribe a uniform local RU architecture. A conventional fully digital or hybrid RU remains attractive when its local aperture is small or moderate, many simultaneous streams are required, or the served user set changes rapidly. In contrast, a tri-hybrid RU adds an antenna or electromagnetic (EM) domain layer after digital and analog signal processing, as illustrated in Fig.~\ref{fig:tri_hybrid_architecture}~\cite{triHybirdHeath,triHybridMimo,heath2026npj}. The last layer can be realized by a dynamic metasurface antenna (DMA), a fluid antenna system (FAS), a tunable parasitic array, or another reconfigurable antenna structure~\cite{newFluidAntenna,wuFluidAntenna}. In~\cite{triHybridMimo}, the FAS realization is modeled specifically as a port-selectable fluid antenna array (FAA). Although their hardware mechanisms differ, these realizations share the architectural objective of moving part of the spatial processing from power-hungry RF and digital paths to the aperture itself.

\begin{figure*}[t]
    \centering
    \subfloat[]
    {
        \includegraphics[width=0.48\textwidth]{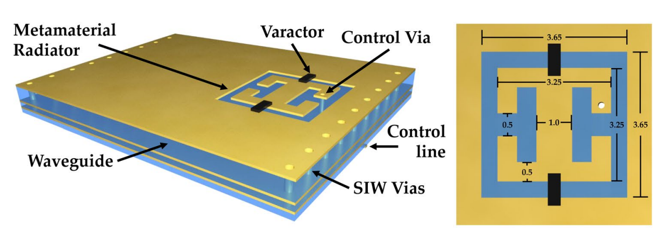}
        \label{dma_nature}
    }
    \hfil
    \subfloat[]
    {
        \includegraphics[width=0.48\textwidth]{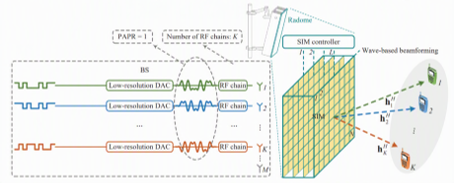}
        \label{sim_ICC}
    }
    \hfil
    \par\vspace{2mm}
    \subfloat[]
    {
        \includegraphics[height=0.30\textwidth]{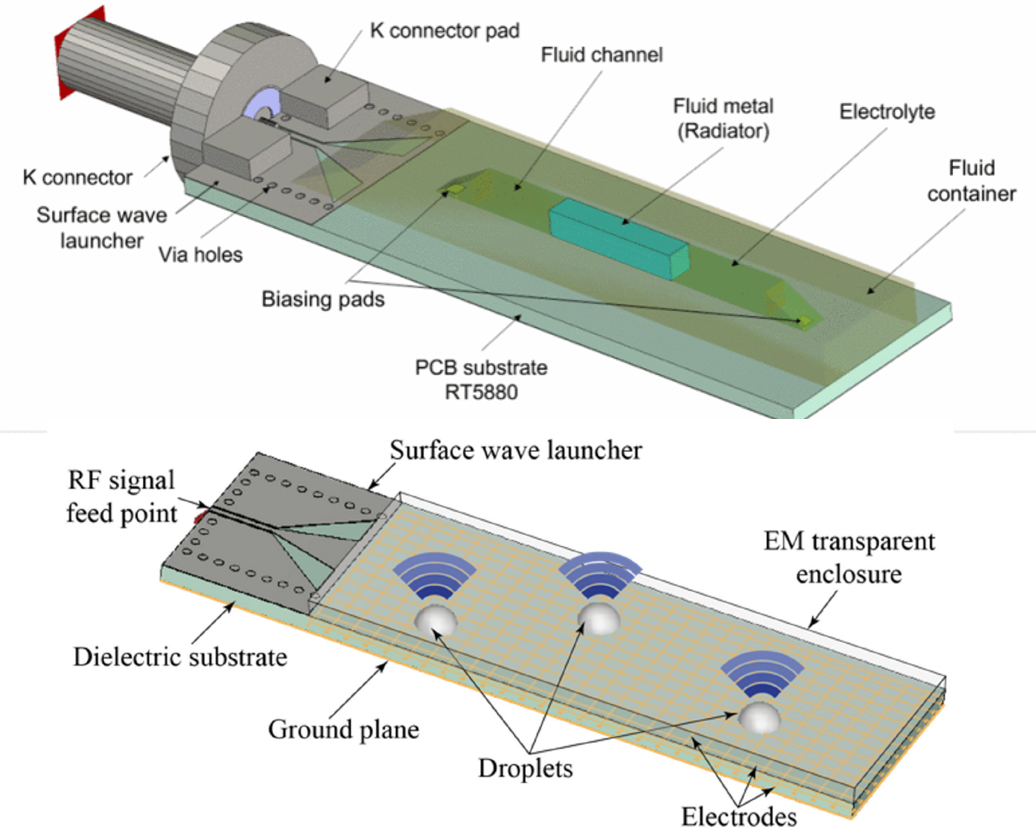}
        \label{fluid_metal}
    }
    \hfil
    \subfloat[]
    {
        \includegraphics[height=0.30\textwidth]{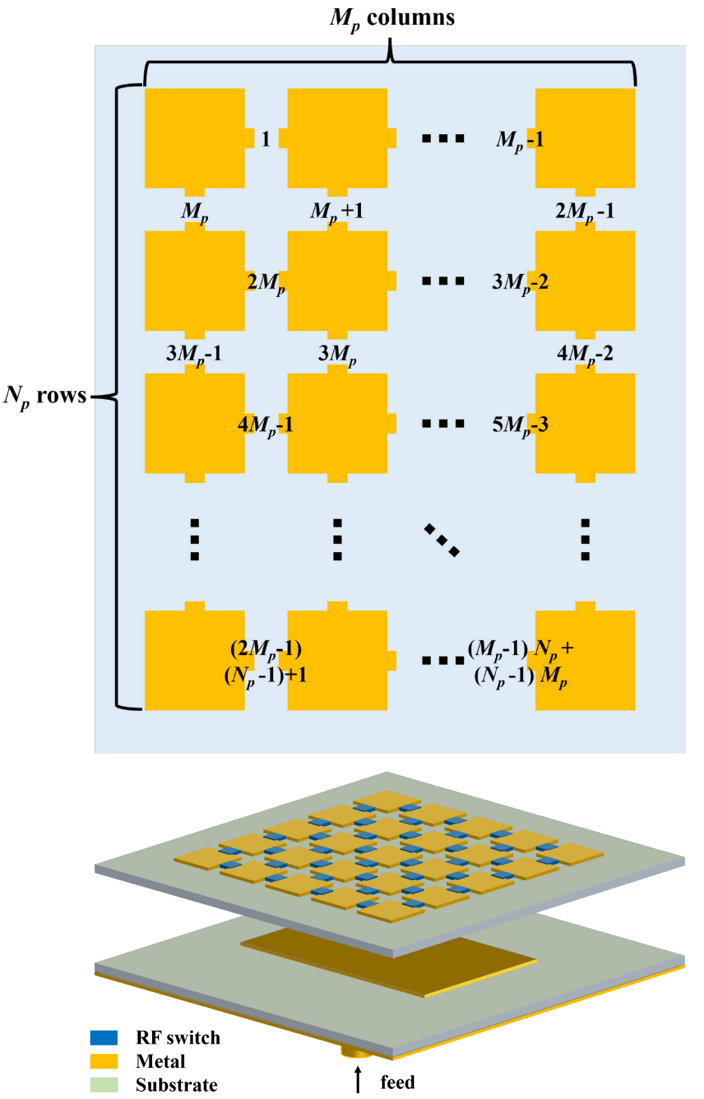}
        \label{ewod}
    }
    \caption{Representative hardware implementations of reconfigurable antennas: (a) Varactor controlled metasurface, adapted from~\cite{boyarskyMetasurfaceAntenna2021}, (b) SIM-based transmitter architecture, adapted from~\cite{anStackedMetasurfaces2023}, (c) Fluid-metal based reconfigurable antenna and electrowetting-on-dielectric (EWOD), adapted from~\cite{wangElectrowettingFluidAntenna2022}, (d) Pixel reconfigurable antenna, adapted from~\cite{newFluidAntenna}.}
    \label{fig:element_designs}
\end{figure*}

\begin{figure}[t]
    \centering
    \subfloat[]
    {
        \includegraphics[width=0.47\textwidth]{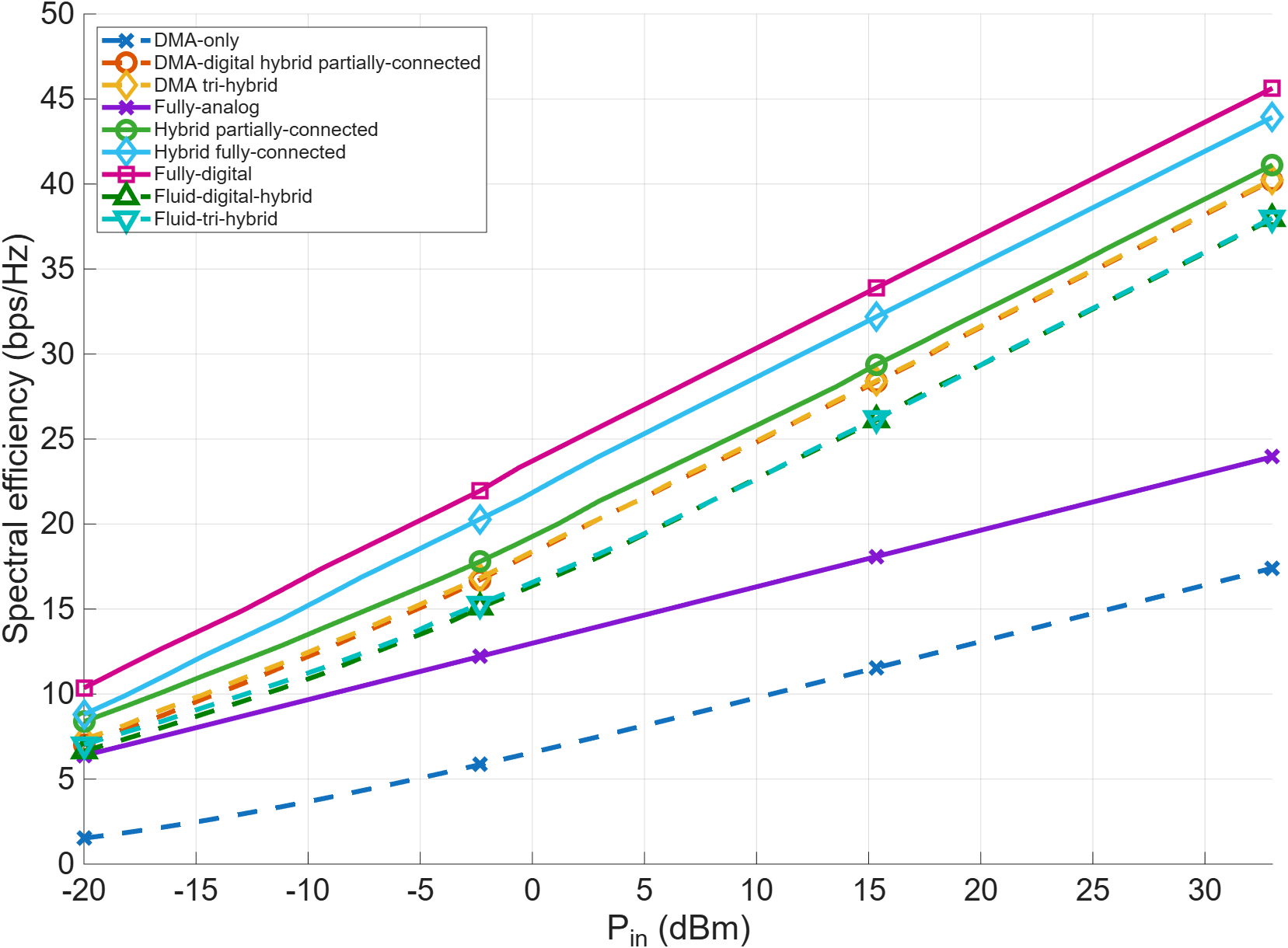}
        \label{cmp_arch_se}
    }
    \hfil
    \subfloat[]
    {
        \includegraphics[width=0.47\textwidth]{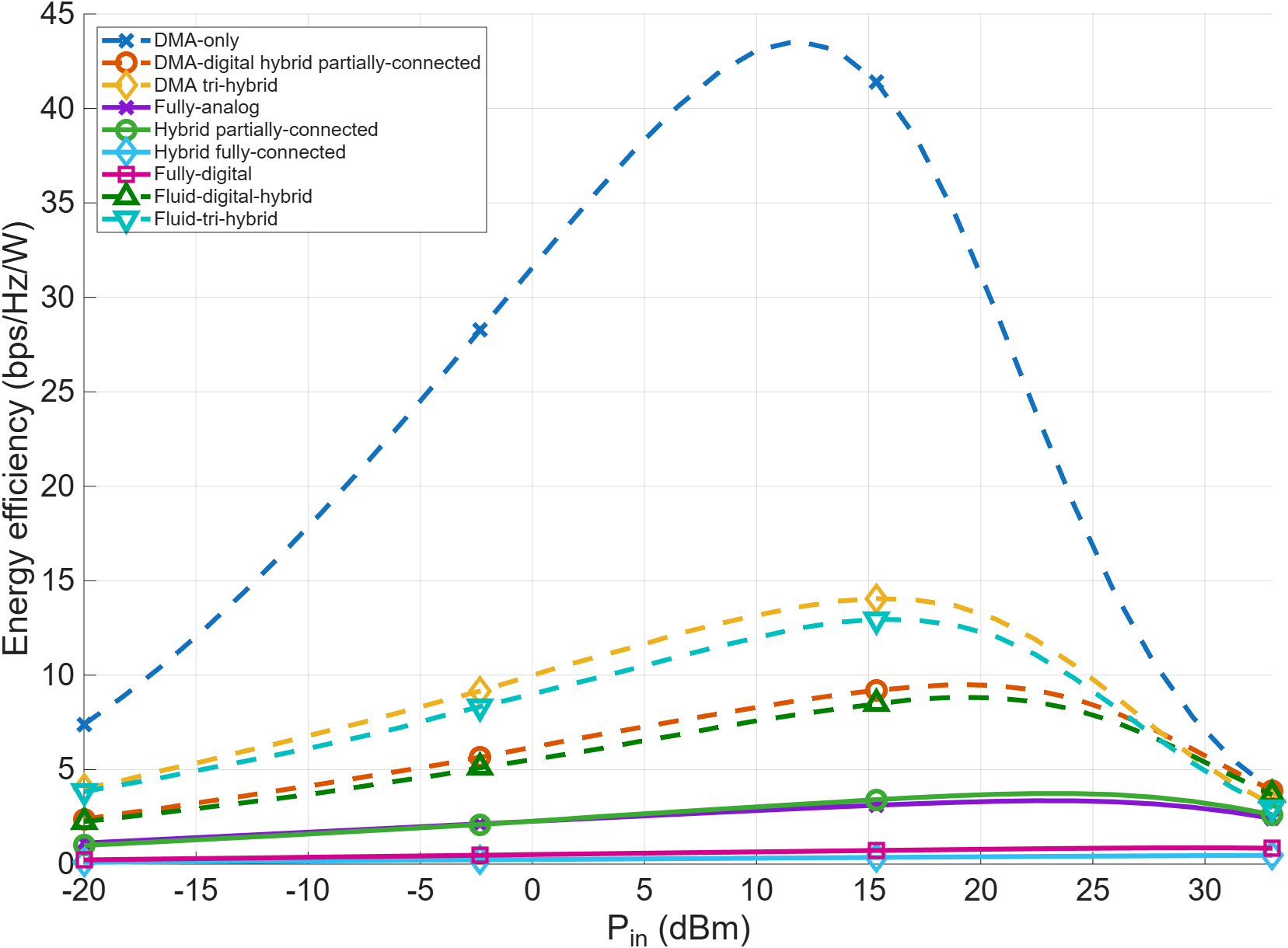}
        \label{cmp_arch_ee}
    }
    \caption{Architecture-level comparison for a 256-radiating-element transmitter. The spectral efficiency result in (a) shows the rate loss incurred by constrained and reconfigurable architectures, whereas the energy efficiency result in (b) shows the corresponding rate-per-watt tradeoff. The comparison illustrates that tri-hybrid and fluid-antenna-based architectures can provide a practical operating point by preserving much of the spectral efficiency while substantially improving energy efficiency.}
    \label{fig:cmp_se_ee}
\end{figure}

Following the tri-hybrid signal model in~\cite{triHybridMimo}, consider
RU \(m\) and subcarrier \(k\). Let \(N_{s,m}\),
\(N_{\mathrm{RF},m}\), \(N_{\mathrm{port},m}\), and
\(N_{\mathrm{rad},m}\) denote the numbers of transmitted data
streams, active RF chains, actively driven antenna/feed ports,
and physical radiating elements at RU \(m\), respectively.
The transmitted data-symbol vector is
\(\mathbf{s}_m[k]\in\mathbb{C}^{N_{s,m}}\), and the resulting
radiator-domain excitation vector is
\(\mathbf{x}_{\mathrm{ant},m}[k]
\in\mathbb{C}^{N_{\mathrm{rad},m}}\).
The tri-hybrid transmit signal is
\begin{align}
    \mathbf{x}_{\mathrm{ant},m}[k]
    ={}&\mathbf{F}_{\mathrm{ant},m}[k]
    \mathbf{F}_{\mathrm{ana},m}
    \mathbf{F}_{\mathrm{dig},m}[k]
    \mathbf{s}_{m}[k].
    \label{eq:tri_hybrid_ru_signal}
\end{align}
The digital precoder
\[
\mathbf{F}_{\mathrm{dig},m}[k]
\in
\mathbb{C}^{N_{\mathrm{RF},m}\times N_{s,m}}
\]
maps the \(N_{s,m}\) data streams to the active RF chains.
The analog precoder
\[
\mathbf{F}_{\mathrm{ana},m}
\in
\mathbb{C}^{N_{\mathrm{port},m}\times N_{\mathrm{RF},m}}
\]
maps the RF-chain outputs to the actively driven antenna/feed
ports. Finally, the antenna-domain mapping
\[
\mathbf{F}_{\mathrm{ant},m}[k]
\in
\mathbb{C}^{N_{\mathrm{rad},m}\times N_{\mathrm{port},m}}
\]
maps the port excitations to the physical radiating aperture.
This separation distinguishes the number of accessible
radiators from the numbers of actively driven ports and RF
chains. 
\[
N_{\mathrm{rad},m}
>
N_{\mathrm{port},m}
\geq
N_{\mathrm{RF},m}
\geq
N_{s,m},
\]
although the exact relationships depend on the local RU
architecture. Consequently, the accessible radiating aperture
can grow without requiring a proportional increase in active
RF chains, data converters, or phase-shifter paths. As illustrated in Fig.~\ref{fig:cmp_se_ee}, representative DMA- and FAS-based tri-hybrid designs provide a favorable spectral efficiency--energy efficiency tradeoff, retaining much of the spectral efficiency of fully digital and conventional hybrid baselines while achieving substantially higher energy efficiency.

The matrix $\mathbf{F}_{\mathrm{ant},m}[k]$ is therefore an architecture-level abstraction rather than a model of one specific antenna. In a DMA, it represents the constrained EM weights that couple a small number of feeds to many subwavelength radiators. These weights can shape the current distribution and radiation pattern with low-power tuning devices, but may obey coupled amplitude and phase constraints and experience waveguide attenuation. In an FAS, the same matrix represents port or position selection over a denser set of candidate radiating locations. Only favorable ports are connected to the available active paths, producing a binary or otherwise discrete mapping rather than the constrained complex weighting of a DMA. FAS thus follows the same low-power scaling principle as DMA: the accessible aperture grows without connecting every candidate radiator to an RF chain. Polarization-reconfigurable and tunable parasitic arrays provide further realizations with their own feasible sets. The common benefit is antenna-domain spatial processing with limited active hardware; the common limitation is a smaller feasible set than independent RF or digital weighting~\cite{triHybridMimo}.

This tradeoff makes tri-hybrid operation conditional rather than universal. DMA-only or FAS/FAA-only precoding architectures can minimize the active hardware, but may incur a pronounced loss of spectral efficiency because they omit one or both of the digital and analog precoding layers and retain tightly constrained antenna-domain precoders. Combining the DMA or FAS layer with both analog and digital precoding provides a more balanced operating point: it preserves more spatial-multiplexing freedom while avoiding the one-RF-chain-per-radiator scaling of a fully digital array~\cite{triHybridMimo}. In DMIMO, this points to a specific role for DMA- or FAS-based tri-hybrid RUs at selected sites where a large local aperture has a persistent function but the required stream rank is limited. Examples include an RU that illuminates a blocked street, supports a cell-edge cluster, maintains a narrow coverage corridor, or provides a high-resolution sensing view. Compact RUs that mainly provide proximity, and hotspot RUs that require many independently updated multiuser streams, may remain conventional. An ultra-dense deployment of tiny RUs can likewise leave too little local aperture for the third processing layer to justify its control, switching, and calibration overhead.

We now model the power consumption of a DMIMO network that mixes tri-hybrid and conventional RUs and activates them selectively. 
Let \(P_{\mathrm{net}}\) denote the total network power
consumption and \(\mathcal{S}\) the set of active RUs. The
architecture indicator of RU \(m\) is
\(\chi_m\in\{\mathrm{C},\mathrm{TH}\}\), where
\(\mathrm{C}\) and \(\mathrm{TH}\) denote conventional and
tri-hybrid hardware, respectively. Let \(q_m\) denote the
number of active RF paths at RU \(m\). Building on the network-level power accounting in~\cite{ngoEnergyEfficiency2018,basharQuantization2019} and the architecture-dependent component analysis in~\cite{triHybridMimo}, we introduce the following implementation-neutral accounting model:
\begin{align}
    P_{\mathrm{net}}
    ={}&\sum_{m\in\mathcal{S}}
    \left[
    \frac{P_{\mathrm{rad},m}}{\eta_{\mathrm{tx},m}^{(\chi_m)}}
    +P_{0,m}
    +P_{\mathrm{RF},m}^{(\chi_m)}(q_m)
    +P_{\mathrm{EM},m}^{(\chi_m)}
    \right]
    \notag\\
    &+P_{\mathrm{FH}}(\mathcal{S})
    +P_{\mathrm{BB}}(\mathcal{S})
    +P_{\mathrm{sync}}(\mathcal{S}).
    \label{eq:dmimo_network_power}
\end{align}
In~\eqref{eq:dmimo_network_power}, \(P_{\mathrm{rad},m}\) is the radiated power of RU
\(m\), and
\(\eta_{\mathrm{tx},m}^{(\chi_m)}\) is its architecture-dependent
end-to-end transmit efficiency, including the PA, insertion,
and radiation efficiencies. The terms
\(P_{0,m}^{(\chi_m)}\),
\(P_{\mathrm{RF},m}^{(\chi_m)}(q_m)\), and
\(P_{\mathrm{EM},m}^{(\chi_m)}\) denote the baseline RU power,
the power consumed by the \(q_m\) active RF paths, and the
antenna/EM-domain control power, respectively. The network-level
terms \(P_{\mathrm{FH}}(\mathcal S)\),
\(P_{\mathrm{BB}}(\mathcal S)\), and
\(P_{\mathrm{sync}}(\mathcal S)\) account for fronthaul
transport, baseband processing, and inter-RU synchronization
over the active-RU set.

The terms $P_{0,m}$, $P_{\mathrm{RF},m}^{(\chi_m)}(r_m)$, and $P_{\mathrm{EM},m}^{(\chi_m)}$ account for the RU's baseline power, active RF paths, and reconfigurable-antenna control, respectively. For a DMA, the last term includes the tuning and control network; for an FAS, it includes port or position selection and the associated switching network. The remaining terms represent fronthaul transport, baseband processing, and synchronization across the active RU set. For a conventional RU, $P_{\mathrm{EM},m}^{(\mathrm{C})}=0$; a tri-hybrid RU adds antenna-control power but may reduce the required number of RF paths for a given accessible aperture. It is energy beneficial only when the resulting reduction in PA and RF-chain power exceeds the added antenna-control, loss, and coordination costs under the same coverage, rate, reliability, latency, and sensing requirements. Equation~\eqref{eq:dmimo_network_power} therefore supports architecture selection rather than a presumption that tri-hybrid hardware is always preferable.

\subsection{Hierarchical Processing and Effective Channel Construction}

Distributing the aperture can reduce PA output power while moving the scaling bottleneck toward fronthaul and baseband processing. A direct extension of centralized MIMO would transport high-rate samples from every candidate RU and construct a full-dimensional channel over all radiating elements. Such an approach retains inactive dimensions in channel acquisition, data conversion, transport, and precoding, even though only a few RUs may dominate a given UE's link~\cite{basharQuantization2019}. The processing architecture should instead follow the same sparsity as RU activation. Slowly varying path loss, blockage, spatial statistics, and traffic demand can identify candidate RU clusters and suitable local antenna precoders. Fast channel acquisition and digital precoding can then be confined to the effective dimensions exposed by the active cluster, leaving detailed CSI methods to Section~VI.

This hierarchy requires a clear interface between local aperture formation and distributed coordination. Each RU should use local analog and antenna-domain processing to map a limited number of active RF signals onto its physical aperture and expose only the resulting effective channels to the central processor. In a tri-hybrid RU, the central processor can operate on an effective channel that already includes $\mathbf{F}_{\mathrm{ant},m}[k]$ and $\mathbf{F}_{\mathrm{ana},m}$ in~\eqref{eq:tri_hybrid_ru_signal}, rather than acquiring and transporting every element-level channel. This reduction is essential: otherwise, the RF-chain saving can be erased by element-wise training and fronthaul transport. A DMA can estimate the channel after its constrained antenna-domain weights have been applied. For an FAS/FAA, a possible research direction is hierarchical sounding: coarse probing first identifies promising port regions, followed by refined channel construction over only the selected ports~\cite{triHybridMimo}. Local combining, precoding, or quantization can further reduce the transported information, but aggressive local processing also limits the observations available for network-wide interference suppression and coherent cooperation. The functional split should therefore be selected according to fronthaul capacity, computation, coordination latency, and residual interference~\cite{basharQuantization2019}.

A common logical interface is particularly important for heterogeneous RUs. Rather than standardizing one internal circuit topology, the system should describe each RU through coordination-relevant capabilities: $N_{\mathrm{rad}}$, $N_{\mathrm{port}}$, and $N_{\mathrm{RF}}$; supported antenna-domain precoding and its frequency dependence; DMA weighting or FAS port-selection constraints; switching latency; insertion and radiation efficiency; calibration validity; admissible stream rank; power modes; and supported coherence level. These descriptors allow the scheduler to construct a distributed channel only over resources that are active and mutually compatible. They also preserve implementation freedom across conventional and tri-hybrid hardware, making the heterogeneous RU distinction operationally meaningful rather than merely descriptive.

\subsection{Adaptive Distributed Apertures for Coverage and Sensing}

The distributed aperture provides a physical means of adapting the coverage discussed in Section~III. Instead of reshaping every requirement from one macro-site aperture, the network can select different illumination and reception points according to location, blockage, traffic, and protocol state. A common signal may use a small set of widely visible RUs, whereas a UE-specific transmission may use the nearest favorable RU together with only the additional nodes needed for reliability or interference control. This flexibility is useful for temporary hotspots and dedicated networks whose service region changes over time. It must nevertheless remain activation aware: widening the served region by turning on many RUs is not a low-power solution unless the added coverage justifies their static and coordination costs.

ISAC obtains a complementary benefit from the same geometry. Spatially separated RUs provide multiple viewing angles and can form monostatic, bistatic, or multistatic sensing links, reducing blind regions that cannot be resolved by one elevated macro site. Their roles need not be identical. A conventional RU may support flexible waveform generation, reception, and multiuser streams, while a selected tri-hybrid RU uses DMA-based pattern, polarization, or frequency control, or FAS-based port and position selection, to form an effective aperture for directional illumination or angular discrimination~\cite{triHybridMimo}. The network can also separate transmitting and receiving roles across RUs when this is more efficient than simultaneous operation at every node. These modes require inter-RU timing, phase, and location calibration, but the required coherence depends on the sensing function and should not be imposed uniformly on all participating RUs.

The common research problem is to schedule communication links, sensing views, RU power modes, and local antenna-domain processing under one end-to-end energy budget. Installed radiating elements should be treated as a pool of potential spatial resources, not as always-on RF paths. Under this view, DMIMO is the primary network-level mechanism: it creates favorable propagation opportunities through proximity and spatial diversity. Conventional and tri-hybrid architectures are conditional RU-level means of exploiting those opportunities. The resulting 6G E-MIMO system is therefore neither a uniformly dense collection of fully digital RUs nor a network composed exclusively of tri-hybrid panels. It is an architecture-aware distributed system that activates the smallest compatible set of RUs, RF paths, streams, and antenna-domain functions needed for the current communication and sensing objectives.

\section{6G E-MIMO CSI Acquisition}


Fundamentally, E-MIMO scaling goes along with multi-user scaling. Adding transmit antennas to a single-user link improves the beamforming gain, but the number of spatial streams stays capped by the antennas at the UE, which remains small. The throughput gain
of a large array therefore comes from multiplexing across users: the base station serves many of them on the same time-frequency resource, and the cell throughput grows with the number scheduled rather than with what any one terminal can do. The measurements above leave room for this. That is to say, the channel rank holds up under equal-aperture scaling, but the room is only potential until the transmitter knows the channel, i.e., channel state information at the transmitter (CSIT)

Typically, multi-user precoding separates users by steering the interference away from each of them. Without accurate CSIT, inter-user interference cannot be controlled and the gain reduces to that of a single user. In the literature, it is well known that the required accuracy scales with the array size: with quantized feedback, preserving the full multiplexing gain of an $N_t$-antenna downlink requires a per-user feedback rate of $B\propto(N_t-1)\log_2\mathrm{SNR}$ bits~\cite{jindal2006}, while in interference-limited cellular networks, a similar scaling holds~\cite{parkFeedbackRate2016}. 
This ties the gain and its price to the same quantity. The array that multiplies
the throughput also multiplies the dimension to be acquired, and CSI acquisition becomes the binding constraint of E-MIMO. This is the antenna-port scaling mismatch identified in Table~\ref{tab:mimo_scaling_breakpoints}. Throughout this section, \(N_t\) denotes the number of logical transmit dimensions exposed to CSI acquisition, corresponding to the configured antenna ports rather than necessarily to the number of physical radiating elements or active RF chains. The objective is therefore not merely to compress a 256-port CSI report, but to make sounding, estimation, feedback, and prediction scale with the effective channel rank rather than with the physical port count.

At its core, CSI acquisition follows an estimate-then-feedback principle: the
base station sounds the channel with pilots, the UE estimates it, and the estimate is
returned, explicitly through feedback in frequency-division duplex (FDD), or
implicitly through uplink reciprocity in time-division duplex (TDD)~\cite{nr38214,parkFeedbackRate2016,moon2026fr3beam}. 
The 5G NR CSI framework is an engineered instance of this principle, and it
breaks at the E-MIMO scale in three ways. 
First, the overhead grows with the array: covering the cell with narrower beams
takes more pilots, and sustaining the multiplexing gain at 256 ports raises the
per-UE feedback toward thousands of bits, a prohibitive cost~\cite{jindal2006,bjornson2025gmimo}. 
Second, the estimate is weak, since sounding precedes beamforming. As the coverage study above showed, the CSI-RS and
the SRS carry only the coarse directional beam, so the estimation stage takes the
full 7\,GHz path loss at close to per-element SNR. Third, it ages before use, as
the doubled Doppler at 7\,GHz shortens the coherence time while the longer
acquisition cycle delays the report, an effect compounded in FDD by the
uplink--downlink split. 

One way out lies in the FR3 channel itself. As the measurements showed, the FR3 channel tends to concentrate in a few angular modes, and its rank grows more slowly than the port
count, so the channel effectively lives in a subspace far smaller than the array
that carries it. That subspace is captured by the long-term statistics of the channel, above all
its spatial covariance. These are set by the propagation
geometry, thus they hold for seconds while the channel itself turns over in milliseconds. 
Let us first suppose these statistics were available. The dominant directions would then be
fixed in advance, without a sweep, and sounding and feedback confined to the
subspace would scale with its rank rather than with the array. The beam gain
would already be in place at estimation, holding up the accuracy at low SNR.
And the subspace outlives the coherence time, so the estimate would no longer go
stale between acquisition and use. 
This leaves two questions. The first is how to redesign the chain around a known
subspace, so that every stage operates inside it. The second is how to obtain the
statistics to begin with, cheaply and even where nothing has been sounded. The
next three subsections take up the first question, and the last one takes up the
second.

{\color{black}{\subsection{A Design That Reduces the Effective Dimension Together with the Reference-Signal and Report Resources}

In the current CSI acquisition chain, both the CSI-RS and the feedback scale with the number
of antenna ports, because the channel is treated as an arbitrary vector in the
full $N_t$-dimensional space and every dimension must be resolved. 
The FR3 channel, however, does not fill that space. The measurements above found
the array response increasingly correlated and the channel power concentrated in
fewer eigenvalues; written in terms of the spatial covariance
$\mathbf{R}=\mathbb{E}[\mathbf{h}\mathbf{h}^{\mathsf{H}}]
=\mathbf{U}\boldsymbol{\Lambda}\mathbf{U}^{\mathsf{H}}$, with eigenvalues ordered
as $\lambda_1\ge\cdots\ge\lambda_{N_t}$, this says that only $r\ll N_t$ of them
are significant. Let \(r\ll N_t\) denote the selected effective channel rank,
and let
\[
\mathbf U_r
=
[\mathbf u_1,\ldots,\mathbf u_r]
\in\mathbb C^{N_t\times r}
\]
contain the \(r\) dominant eigenvectors of \(\mathbf R\).
The channel can then be approximated as
\begin{align}
  \mathbf{h}\simeq\mathbf{U}_r\,\mathbf{g},\; \mathbf{g}\in\mathbb{C}^{r},
  \label{eq:lowdim}
\end{align}
where $\mathbf{g}$ is the reduced-dimensional channel coefficient. 
For $\mathbf{h}\sim\mathcal{CN}(\mathbf{0},\mathbf{R})$, projecting onto
$\mathbf{U}_r$ is the Karhunen--Lo\`eve transform of the channel. It decorrelates
the coefficients, and no other $r$-dimensional subspace captures more energy than
$\mathbf{U}_r$. The error in \eqref{eq:lowdim} is thus the smallest any dimension $r$ can give.
Acquiring these $r$ coefficients is therefore enough to represent the channel,
and resolving all $N_t$ dimensions is wasteful. 



Suppose the covariance $\mathbf{R}$ of a given UE is known in advance, and with it
the subspace $\mathbf{U}_r$. Acquisition can then be confined to that subspace.
The base station precodes the CSI-RS onto the $r$ dominant eigen-beams
$\mathbf{U}_r$, so $r$ beamformed ports are sounded in place of $N_t$. The UE
estimates and reports only the $r$ coefficients $\mathbf{g}$, so the quantization
is carried out in $\mathbb{C}^{r}$ rather than $\mathbb{C}^{N_t}$ and the payload
follows the reduced dimension. 
Every stage now scales with the effective rank rather than the port
count, and the two quantities scale differently. The rank $r$ is a
space--bandwidth product, set by the aperture measured in wavelengths
together with the angular support of the propagation. The port count
$N_{t}$ is set by how that aperture is sampled and by the RF budget.
In the equal-aperture transition from 3.5 to 7\,GHz, $N_{t}$
quadruples, while the narrowing angular spread of Section II-B offsets
part of the aperture growth, so $r$ grows sublinearly in $N_t$, well below its fourfold increase
Any further growth of the array that samples the same aperture more
densely leaves $r$ unchanged altogether. The saving $N_{t}/r$
therefore widens along the E-MIMO trajectory rather than staying
fixed, and the acquisition cost follows the propagation rather than
the hardware. This distinction is sharpest for the architectures of Section V. A
tri-hybrid RU satisfies $N_{\mathrm{rad}} > N_{\mathrm{port}} \geq
N_{\mathrm{RF}}$, enlarging the radiating aperture without scaling the
active chains, and the channel it exposes still has rank $r$. The
acquisition cost thus follows neither the radiator count nor the port
count, which is the sense in which it is decoupled from the scale of
the array.

A similar approach has been studied before. JSDM~\cite{adhikary2013jsdm} groups UEs whose
covariances span similar subspaces, maps the array onto the group subspace
with an outer precoder that depends only on second-order statistics, and
carries downlink training and feedback on the resulting effective channel
rather than on the full array. One property of that construction matters
throughout what follows: the UE observes only $\mathbf{g}$ and never needs
$\mathbf{U}_{r}$, so the subspace remains a base-station-side quantity that
costs nothing to convey. What the JSDM architecture does not settle is the rest of the acquisition
chain. Three problems remain at the E-MIMO scale. The channel has to be
estimated at the sounding stage, 
so each port sees only per-element SNR. The estimate then has to survive until it
is used, against a Doppler that doubles at 7\,GHz. And the codebook, the CQI,
and the scheduler have to be rebuilt so that each consumes $r$ dimensions
rather than $N_{t}$.

\subsection{Acquisition That Overcomes Low SNR and Channel Variation}

The first subsection reduced the number of dimensions to be acquired. This one
turns to how reliably each is acquired, against the low SNR and the aging of the
channel. Both difficulties, like the one before, ease once the covariance
$\mathbf{R}$ is known. We consider the low SNR first. The known subspace lets the base
station steer the CSI-RS along the dominant eigen-beams $\mathbf{U}_r$ instead of
sounding every port. The pilot energy then concentrates on the $r$ modes where the channel power
lies, and the estimate sees the array gain that the current framework supplies
only after beamforming. The covariance moves that gain forward into the
estimation stage.

The covariance enters the estimator as well. With the pilots beamformed along \(\mathbf U_r\), and after
pilot despreading and normalization by the pilot energy, the
UE obtains the reduced-dimensional observation
\begin{align}
  \mathbf{y}=\mathbf{g}+\mathbf{n},\; 
  \mathbf{n}\sim\mathcal{CN}(\mathbf{0},\sigma^{2}\mathbf{I}_r),
  \label{eq:obs}
\end{align}
where \(\mathbf y,\mathbf g,\mathbf n\in\mathbb C^r\), \(\sigma^2\) is the effective noise variance after pilot
despreading, and \(\mathbf I_r\) is the \(r\times r\) identity matrix. Within the selected subspace, the covariance of
\(\mathbf g\) is
\[
\boldsymbol\Lambda_r
=
\operatorname{diag}(\lambda_1,\ldots,\lambda_r).
\]
The LMMSE estimate of \(\mathbf g\) is therefore
\begin{align}
  \hat{\mathbf{g}}
  =\boldsymbol{\Lambda}_r\big(\boldsymbol{\Lambda}_r+\sigma^2\mathbf{I}\big)^{-1}
   \,\mathbf{y},
  \label{eq:lmmse}
\end{align}
with $\boldsymbol{\Lambda}_r=\mathrm{diag}(\lambda_1,\dots,\lambda_r)$ the
covariance of $\mathbf{g}$, diagonal because $\mathbf{U}_r$ is the eigenbasis. The filter therefore reduces to a per-mode scalar $\lambda_i/(\lambda_i+\sigma^2)$: strong modes pass through, weak ones are shrunk toward zero, and much of the
accuracy that a least-squares estimate would lose at per-element SNR is recovered. 
This filter can also be learned from the pilot observations alone, without
explicit knowledge of the covariance~\cite{ammse2025}.


Second, the covariance also addresses channel aging. Doubling the carrier doubles
the Doppler, so the channel at 7\,GHz turns over twice as fast as at 3.5\,GHz. 
The subspace does not change the rate of the Doppler, but it changes the form. 
Write the channel at time $t$ as
$\mathbf{h}(t)\simeq\mathbf{U}_r\,\mathbf{g}(t)$. The subspace is set by the
propagation geometry and holds over the prediction horizon, so only
$\mathbf{g}(t)\in\mathbb{C}^{r}$ moves from slot to slot, and predicting
$\mathbf{h}(t+\tau)$ means predicting $\mathbf{g}(t+\tau)$. 
This coefficient, in turn, varies with structure. 
A path's Doppler is set by the direction it arrives
from, so separating directions separates Doppler. Here the E-MIMO array size becomes an advantage: at hundreds of ports the eigen-beams are narrow enough to admit one cluster and reject the others, so each coefficient turns at a single rate. 
Consequently, the subspace does not slow the channel, but it exposes the structure of its
variation, and that structure is what makes extrapolation possible over a longer horizon at the same pilot cost.



}}

\subsection{Overcoming CSI Uncertainty in PMI, Link Adaptation, and Scheduling}

Even with the acquisition above, CSIT is never exact, and the residual
uncertainty incurs performance loss in the precoder choice, the link adaptation, and
the scheduling. 
In this subsection, we discuss how the channel covariance can be used to mitigate
the performance loss caused by the CSI uncertainty.

The PMI reports a precoder by selecting an entry from a codebook, and the
accuracy of that report is governed by the dimension over which the codebook
must spread its entries. Define the normalized channel direction as $\widetilde{\mathbf h}
  =
  {\mathbf h}/{\|\mathbf h\|}$, and let \(\mathcal C_{d}\subset\mathbb C^{d}\) be a unit-norm codebook with
\(2^{B}\) entries, where \(B\) is the feedback payload in bits and \(d\) is the
complex dimension the codebook must span. The UE reports the index of
\[
  \mathbf c^{\star}
  =
  \arg\max_{\mathbf c\in\mathcal C_{d}}
  \big|\widetilde{\mathbf h}^{\mathsf H}\mathbf c\big|^{2},
\]
and the resulting beamforming loss is $D
  =
  \mathbb E\big[\,
  1-\big|\widetilde{\mathbf h}^{\mathsf H}\mathbf c^{\star}\big|^{2}
  \big]$. Quantizing a unit-norm direction is quantization on the complex projective
space \(\mathbb{CP}^{d-1}\), a manifold of real dimension \(2(d-1)\) on which a
ball of chordal radius \(\delta\) occupies a volume fraction of order
\(\delta^{2(d-1)}\). Covering the space with \(2^{B}\) such balls therefore
requires \(2^{B}\delta^{2(d-1)}\gtrsim 1\), so that the beamforming loss is obtained as 
\begin{align}
  D\;\approx\;2^{-\frac{B}{d-1}},
  \label{eq:covering}
\end{align}
and each additional bit buys \(3/(d-1)\)\,dB of accuracy. Random vector
quantization attains the same exponent up to a constant
factor~\cite{jindal2006}. \eqref{eq:covering} is thus not an artifact
of one particular construction but the best scaling available to any codebook
that must span \(d\) dimensions, and the design question is what \(d\) that is.


Absent statistical side information, the codebook cannot be tailored to any
particular UE and must instead be matched to the distribution of
\(\widetilde{\mathbf h}\) marginalized over every channel the cell may present.
That marginal is spread over the full angular support the array can resolve, so
every direction in \(\mathbb{CP}^{N_{t}-1}\) has to be represented, giving
\(d=N_{t}\) and $D_{\mathrm{full}}\approx 2^{-\frac{B}{N_{t}-1}}$. 
For this reason, at \(N_{t}=256\) each bit is then worth about \(0.01\)\,dB, and doubling the
array halves even that. The payload is spread evenly over all \(N_{t}\)
directions while any given channel occupies only \(r\) of them, so almost all of
it is spent resolving directions that channel never uses.

The subspace removes this waste without new codebook machinery. Conditioned on
$\mathbf{R}$, the direction to be reported is confined to the span of
$\mathbf{U}_{r}$ rather than ranging over $\mathbb{CP}^{N_{t}-1}$, and the
acquisition chain above already delivers it in that form: with the CSI-RS
precoded onto $\mathbf{U}_{r}$, what the UE observes and reports is the
effective channel $\mathbf{g}\in\mathbb{C}^{r}$ of \eqref{eq:lowdim}, not
$\mathbf{h}$. Writing $\widetilde{\mathbf g}=\mathbf g/\|\mathbf g\|$, the
report is an ordinary $r$-dimensional quantization,
\begin{align}
  \mathbf c^{\star}=\arg\max_{\mathbf c\in\mathcal C_{r}}
  \big|\widetilde{\mathbf g}^{\mathsf H}\mathbf c\big|^{2},\;
  \hat{\mathbf w}=\mathbf U_{r}\,\mathbf c^{\star},
  \label{eq:pmi_eig}
\end{align}
where $\mathcal C_{r}\subset\mathbb C^{r}$ is a unit-norm codebook with $2^{B}$
entries, fixed at both ends in advance exactly as $\mathcal C_{N_{t}}$ was. 
We note that the UE does not need $\mathbf U_{r}$ itself, since the base station applies it when
reconstructing the precoder; the subspace stays a network-side quantity and
consumes none of the payload. What changes is not how the codebook is built but
the dimension it is built for, and the covariance is what places that dimension
where the channel lives.

Writing $\varepsilon_{r}=\sum_{i>r}\lambda_{i}/\sum_{i}\lambda_{i}$ for the
fraction of channel energy left outside the subspace, the loss becomes
\begin{align}
  D_{\mathrm{eig}}\approx \varepsilon_{r}
  +(1-\varepsilon_{r})\,2^{-\frac{B}{r-1}} .
  \label{eq:deig}
\end{align}
The exponent is now set by $r$: each bit buys $3/(r-1)$\,dB, which at
$N_{t}=256$ and $r=8$ raises the return from $0.01$ to $0.43$\,dB, a factor of
$(N_{t}-1)/(r-1)\approx36$. Read the other way, holding the accuracy fixed cuts
the payload by the same factor. The estimate is conservative, since
$\mathbf{g}$ is not isotropic and its remaining eigenvalue spread can only help
the quantizer. The first term is the price of the reduction: \eqref{eq:deig} does not vanish with $B$ but floors at
$\varepsilon_{r}$, so bits beyond roughly $(r-1)\log_{2}(1/\varepsilon_{r})$ are
wasted. The rank should therefore be chosen so that the floor sits below the
accuracy the link needs, which at operating SNR $\rho$ means
$\varepsilon_{r}\lesssim1/\rho$. 
The angular concentration reported in Section~II-B is what allows this to be met at a rank far below the port count.

The idea of concentrating the payload has partial precedent in NR. In the NR Type-II codebook, \(L\) denotes the number of selected beams from an oversampled DFT grid. The UE reports a linear combination of these \(L\) beams rather than a full-dimensional channel direction. What it does not do is remove the search: the beams are found
anew from an instantaneous measurement, their indices ride in every report, the
choice is confined to the grid, and $L$ stays small to keep the payload
bounded. Held as a statistic instead, the subspace is known before the report
is requested, so it costs no indices, is not tied to a grid, and can be as wide
as the propagation warrants rather than as narrow as the payload permits. A
Type~II style combination remains the natural way to spend the bits once the
space they cover has been fixed.

An accurate CQI is harder to obtain. In MU-MIMO, a UE computes its CQI from its own channel and its
own precoder, but the base station then co-schedules it with others whose
precoders behave as interference. That decision is made after the report, so the
reported SINR and the obtained SINR are different quantities. When the CQI is
wrong, the link adaptation fails in one of two ways: an optimistic report leads
to decoding failures and repeated retransmissions, while a pessimistic one wastes
the link by sending fewer bits than the channel could carry.


For a target UE \(k\), let
\(\mathbf h_k\in\mathbb C^{N_t}\) denote its channel and
\[
\mathbf R_k
=
\mathbb E[\mathbf h_k\mathbf h_k^H]
\]
its long-term channel covariance. Let
\(\widehat{\mathbf h}_k\) denote the channel estimate,
\(\mathbf e_k=\mathbf h_k-\widehat{\mathbf h}_k\) the
estimation error, and
\[
\mathbf R_{e,k}
=
\mathbb E[\mathbf e_k\mathbf e_k^H]
\]
the corresponding error covariance. For a co-scheduled UE
\(j\), let \(\mathbf w_j\) denote its transmit precoder. This mismatch can be alleviated if the base station holds the long-term
statistics $\mathbf{R}_{k}$ of each UE. From $\mathbf{R}_{k}$ and the
acquisition chain above, the base station can form the error covariance
$\mathbf{R}_{\mathbf{e},k}$ of its own estimate of $\mathbf{h}_{k}$. This is
the quantity that matters for CQI. A multiuser precoder is built to suppress
its co-scheduled UEs, so what leaks into UE $k$ is not its channel but the part
of it the estimate missed. The expected residual interference power contributed by the precoder of UE \(j\) to UE \(k\) is therefore $\mathbf{w}_{j}^{\mathsf{H}}\mathbf{R}_{\mathbf{e},k}\mathbf{w}_{j}$,
which the base station can evaluate before transmitting. The gap between the reported and the obtained SINR
narrows accordingly.

The covariance is also very useful in the scheduling decision itself.
Since $\mathbf{w}_j^{\mathsf{H}}\mathbf{R}_k\mathbf{w}_j$ is small when the
dominant subspaces of UEs $j$ and $k$ are nearly orthogonal, the base station can
group UEs by their subspaces and keep the interference small to begin with. 
The CQI mismatch is then not merely corrected but further reduced by this construction. 
And if the error covariance of the estimate is carried along as well, the SINR becomes a
distribution rather than a point, so the rate can be chosen to meet an outage
target.

\begin{figure}[t]
    \centering
    \includegraphics[width=\columnwidth]{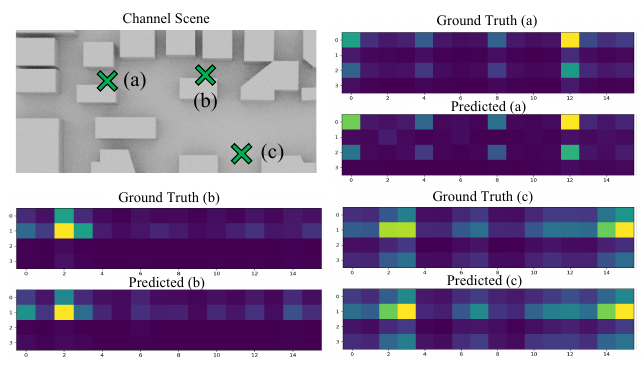}
    \caption{Comparison of the ground-truth and rendered beamspace channel statistics at three unseen UE locations. The selected locations are marked in the channel scene, and the corresponding ground-truth and predicted statistics are shown in (a)--(c).}
    \label{fig:rendering}
\end{figure}


\subsection{Acquiring Long-Term Statistics at Unseen UE Locations}

The preceding subsections all assume that the statistics are already at the
base station. This subsection discusses how they can be obtained efficiently. 
The conventional answer is to average over observations of the UE itself. The base station
sounds, accumulates, and refines its estimate of $\mathbf{R}_{k}$ as the
connection proceeds. This approach works, but it indexes the statistics by the terminal
rather than by the environment. The spatial structure being measured belongs to
the location. The record of it belongs to the UE, and is discarded when the UE
leaves. A newly arriving terminal therefore should start from nothing, however many
others have passed through the same place before it. It must build its
statistics out of the very sounding they were meant to reduce, and it has none
of them at initial access, where Section~III showed the link budget to be
worst.

These statistics are set by the propagation geometry, and geometry is a
property of place. Two terminals standing in the same spot face the same
buildings and the same scatterers. They see the same dominant directions, the
same distribution of power across them, and the same blockage. The natural
index for the record is therefore the location, not the terminal. Re-indexing
this way changes what an observation is worth. Under a per-UE index, it serves
one connection and expires with it. Under a location index, it joins a record
that accumulates over the deployment. It also lets observations speak for
places where none were made, since one scene produces the statistics everywhere
in a cell. That scene can be learned from radio observations gathered wherever
terminals happened to be, and then queried at any position, including positions
never visited.


Accordingly, a promising direction is to render these statistics from a model
of the environment rather than accumulate them from the terminal. The base
station learns a channel scene representation from radio observations collected
across the deployment, and queries it at a target location $\mathbf{p}$ to
render the long-term statistics there, without sounding that location to build them.

\begin{figure}[t]
    \centering
    \includegraphics[width=\columnwidth]{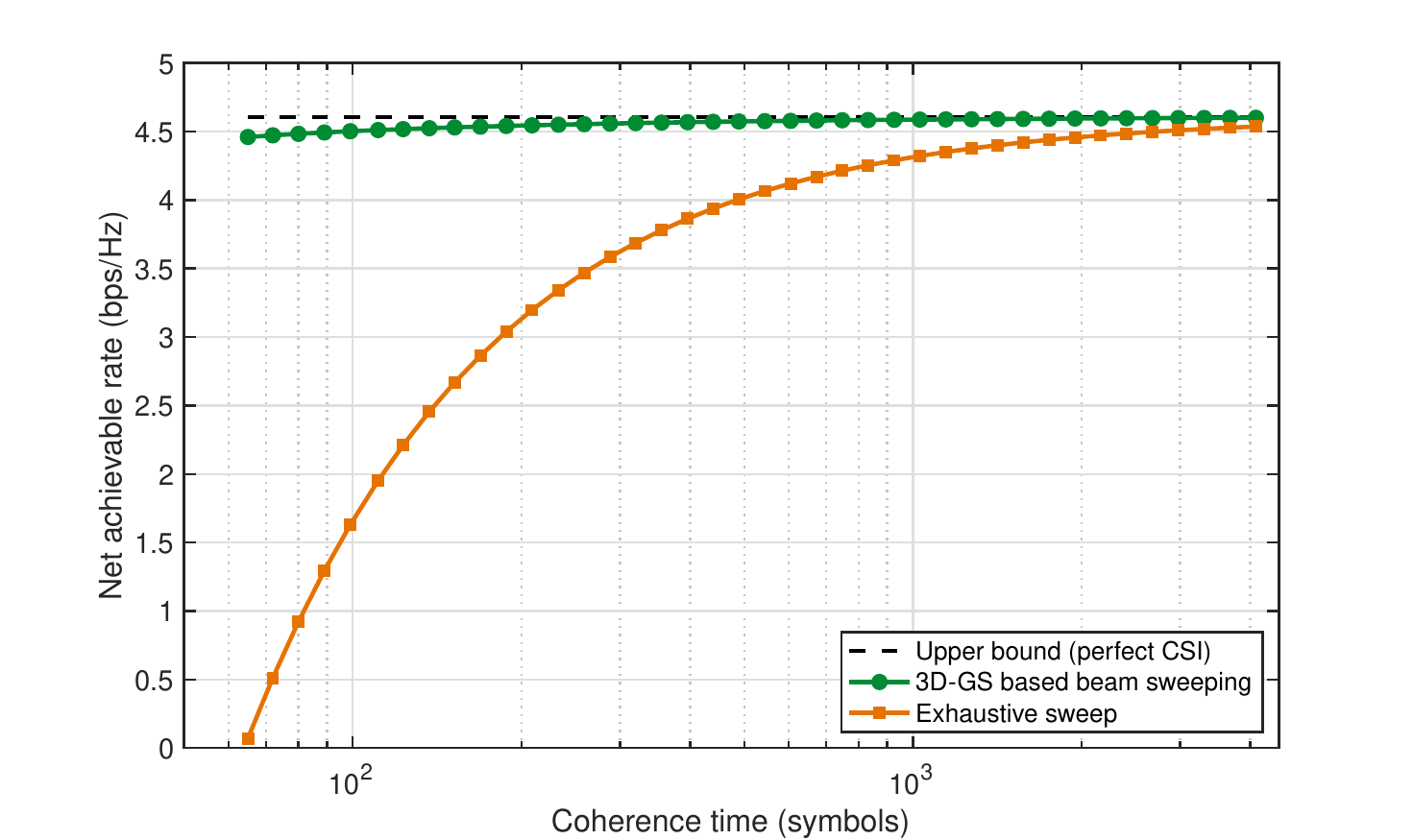}
    \caption{Net achievable rate versus coherence-block length. The 3D-GS based beam sweeping approaches the upper bound with no overhead, achieving substantially lower overhead than exhaustive beam sweep.}
    \label{fig:3dgs_netse}
\end{figure}

As a preliminary step toward this direction, we consider rendering the
beamspace power profile, the beam directions along which the channel power
concentrates. While this is not the exact covariance, its diagonal can capture the power across beam pairs without the phase relations among them. 
In the receive and transmit DFT bases, we have 

Let \(\mathbf p\) denote the UE location and \(t\) the
channel-observation time. The MIMO channel between \(N_t\)
transmit dimensions and \(N_r\) receive dimensions is denoted
by
\begin{align}
    \mathbf{H}_{b}(\mathbf{p},t)=\mathbf{A}_{r}^{\sf H}\,
    \mathbf{H}(\mathbf{p},t)\,\mathbf{A}_{t},
\end{align}
where 
\(\mathbf A_r\in\mathbb C^{N_r\times N_r}\) and
\(\mathbf A_t\in\mathbb C^{N_t\times N_t}\)
denote the unitary receive- and transmit-side DFT matrices,
respectively. The resulting beamspace channel is
\begin{align}
    [\hat{\mathbf{S}}(\mathbf{p})]_{m,n}\approx\mathbb{E}_{t}
     \left[\big|[\mathbf{H}_{b}(\mathbf{p},t)]_{m,n}\big|^{2}\right],
    \label{eq:rendered}
\end{align}
i.e., the channel power on receive beam $m$ and transmit beam $n$. 
On a large uniform array, the spatial covariance is asymptotically diagonalized
by the DFT~\cite{adhikary2013jsdm}, so its dominant eigenvectors align with the
strongest DFT beams. The heavy entries of $\hat{\mathbf{S}}(\mathbf{p})$
therefore mark the support on which $\mathbf{U}_{r}$ concentrates.
The E-MIMO regime is where this approximation is tightest.

We now state the rendering problem the profile poses. The input is a set of
beamspace observations $\{\hat{\mathbf{S}}(\mathbf{p}_{i})\}$ measured at
locations $\mathbf{p}_{i}$ that terminals (i.e., UEs) have visited. The output is $\hat{\mathbf{S}}(\mathbf{p})$ at a location $\mathbf{p}$ that was never observed, so the task is not to denoise or average the observations but to
generalize across space between them.
This is the main point that distinguishes rendering from ordinary covariance
estimation: the statistic is required exactly where no measurement exists, therefore the implicit mapping function from location to statistic must be learned from the surrounding ones. 
The propagation geometry is what makes this well posed, since the same scene generates the profile at every point in the cell, and observations at visited locations constrain that scene rather than only their own points.

This channel rendering problem is deeply connected to novel-view
synthesis \cite{mildenhall2020nerf, kerbl:acm:23} from computer vision, 
which has recently been brought into the RF domain to reconstruct the channel from sparse measurements~\cite{zhao2023nerf2, wen2025wrfgs}. 
Different from prior work that reconstructs the instantaneous channel, our approach reads a UE location as the viewpoint and its E-MIMO beamspace profile as the view, whose dimension is set by the large port count of the E-MIMO array. 
To address this, we construct a renderer that adapts the 3D-GS technique~\cite{kerbl:acm:23} to the E-MIMO beamspace. The scene is represented by a set of Gaussian primitives fitted to beamspace observations gathered at
known UE locations, and querying the representation at an unseen location
$\mathbf{p}$ renders the beam profile $\hat{\mathbf{S}}(\mathbf{p})$ there. In
contrast to the RF renderers above, which produce the channel of a single link, the primitives here are splatted onto the transmit--receive beam grid, thus a single query returns the full $N_{r}N_{t}$ profile rather than a per-link
quantity. 

Fig.~\ref{fig:rendering} shows the rendered profile against the ground truth at three unseen locations. 
As shown in the figure, our renderer places the power on the same few beam pairs that the true profile occupies, even where no measurement was taken. 
Such rendered channel information can be used throughout the CSI acquisition
chain of the preceding subsections. Because the profile marks where the power
concentrates, out of the full $N_{r}N_{t}$ beam grid, it identifies the support
on which the subspace $\mathbf{U}_{r}$ lives before any location-specific pilot
is sent. The base station can then confine its CSI-RS and feedback to that
support, and recover the exact $\mathbf{U}_{r}$ and its eigenvalues by sounding
only the flagged pairs. 
That said, the rendered profile does not replace channel sounding, since it captures where the channel power concentrates but not the finer structure needed to form $\mathbf{U}_{r}$ exactly.
Its value is instead in making that sounding scalable. By fixing the support in advance, from a scene learned once
rather than from pilots spent per UE, it decouples the cost of acquisition from the port count and lets the reference signals, the feedback, and the estimate all scale with the effective rank rather than with the size of the E-MIMO
array.





Fig.~\ref{fig:3dgs_netse} plots the net achievable rate against the coherence-block length $T_{\mathrm{c}}$, measured in channel uses. A smaller sweeping overhead leaves
more symbols for data, so the net rate approaches the ideal rate that no
acquisition overhead would give. With the rendered beamspace statistic, the 3D-GS based method reaches close to this upper bound even at short coherence lengths. A conventional sweep, by contrast, searches the beam pairs at every dwell. This leads to a much larger $T_{\mathrm{c}}$ to amortize that cost, whether it tests all
$N_{r}N_{t}$ pairs or narrows them through the hierarchical search of
Section~III. 


Rendering the channel statistics from a model of the environment is, in the
end, a core technique to acquiring CSI at the E-MIMO scale. It fixes the beam support
in advance, from a scene learned once rather than from pilots spent per UE, so
the reference signals, the feedback, and the estimate scale with the effective rank rather than the port count. What we have shown here is one step of this: rendering the beamspace power profile is enough to point the sounding
and cut its overhead. Rendering the full channel covariance, from which the
subspace $\mathbf{U}_{r}$ would follow directly without any sounding at all,
remains open, and is a natural direction for turning this approach into a
complete CSI-acquisition framework for E-MIMO.




\section{Conclusions}
The upper-mid band has emerged as a leading spectrum candidate for wide-area 6G deployment, making fixed-aperture scaling to hundreds of antenna ports, E-MIMO, a central direction of 6G physical-layer research. The equal-aperture argument indicates that the higher free-space path loss at 7~GHz can, in principle, be compensated while reusing the physical aperture and sites of existing 5G networks. Translating this potential into a deployable system, however, reveals four coupled breakpoints: maintaining effective coverage across physical channels and protocol states; implementing wideband, energy-efficient RF hardware; containing the power consumption of a large radiating aperture that would otherwise require hundreds of active RF chains; and acquiring sufficiently accurate CSI with manageable sounding and feedback overhead.

This paper has examined these breakpoints, explained why a uniform extension of the 5G NR architecture becomes inefficient or infeasible at the E-MIMO scale, and identified representative research directions. A common principle connects the four: the number of radiating elements, antenna ports, active RF chains, and acquired channel dimensions need not---and should not---scale at the same rate. The accessible aperture may grow to hundreds or thousands of radiating elements, whereas active hardware and protocol overhead should scale primarily with the spatial degrees of freedom required by the propagation environment, traffic, and service objectives. Realizing this principle requires protocol-aware coverage enhancement, wideband RF--antenna--baseband co-design, architecture-aware activation of co-located and distributed apertures, and CSI acquisition that exploits the long-term structure and effective rank of the channel. These directions move E-MIMO beyond a uniform enlargement of 5G massive MIMO toward an adaptive and deployable 6G architecture. Their ultimate realization will depend on the continuing standardization of upper-mid-band operation and on joint advances in propagation characterization, hardware, array architecture, and signal processing.

\section*{Acknowledgement}
The authors would like to thank Minji Phi and Juneyoung Park for their valuable help in the simulations, experiments, and data collection.

\bibliographystyle{IEEEtran}
\bibliography{ref}
\vfill

\end{document}